\documentclass[useAMS,usenatbib]{mn2e}
\usepackage{graphicx}

\usepackage{caption}
\usepackage{amsmath}
\usepackage{natbib}
\usepackage{color}
\usepackage[english]{babel}

\def \ascl Astrophysics Source Code Library
\def \pasj{PASJ}

\newcommand{\noop}[1]{}

\title[Shape evolution in DM halos]{Haloes at the ragged edge: The importance of the splashback radius }
\author[Snaith et al.]{O. N. Snaith$^{1,2}$, J. Bailin$^{2}$, A. Knebe$^{3,4}$,
 G. Stinson$^5$, J.  Wadsley$^6$, H. Couchman$^6$\\
$^1$ School of Physics, Korea Institute for Advanced Study, 85 Hoegiro, Dongdaemungu, Seoul 02455, Republic of Korea\\
$^2$ Department of Physics and Astronomy, University of Alabama, Box 870324, Tuscaloosa, AL 35487-0324, USA\\
$^3$ Departamento de F\'isica Te\'orica, M\'odulo , Universidad Aut\'onomade Madrid, Cantoblanco, E-28049, Spain \\
$^4$ Astro-UAM, UAM, Unidad Asociada CSIC\\
$^5$ Max-Planck-Institut fur Astronomie, Konigstuhl 17, D-69117, Heidelberg, Germany \\
$^6$ Department of Physics and Astronomy, McMaster University, Hamilton, Ontario, L8S 4M1, Canada \\
}
\date{\today}
\begin{document}

\maketitle

\begin{abstract}
We have explored the outskirts of dark matter haloes out to 2.5 times the virial radius using a large sample of halos drawn from Illustris, along with a set of zoom simulations (MUGS). Using these, we make a systematic exploration of the shape profile beyond R$_{vir}$. In the mean sphericity profile of Illustris halos we identify a dip close to the virial radius, which is robust across a broad range of masses and infall rates. The inner edge of this feature may be related to the virial radius and the outer edge with the splashback radius. Due to the high halo-to-halo variation this result is visible only on average. However, in four individual halos in the MUGS sample, a decrease in the sphericity and a subsequent recovery is evident close to the splashback radius. We find that this feature persists for several Gyr, growing with the halo. This feature appears at the interface between the spherical halo density distribution and the filamentary structure in the environment. The shape feature is strongest when there is a high rate of infall, implying that the effect is due to the mixing of accreting and virializing material. The filamentary velocity field becomes rapidly mixed in the halo region inside the virial radius, with the area between this and the splashback radius serving as the transition region. We also identify a long-lasting and smoothly evolving splashback region in the radial density gradient in many of the MUGS halos.
\end{abstract}

\section{Introduction}

The shape of dark matter halos contains considerable information about the properties of the halo beyond the naive assumption of spherical symmetry. It is common to describe the shape of dark matter halos in terms of the eigenvectors of the second moment tensor, which provide axial ratios for the shape ellipsoid. Contrary to the assumption of spherical infall of material, halos accrete matter from cosmic filaments along preferential directions \citep[e.g.][]{Knebe2004,Libeskind2011,Libeskind2014}. \citet{Frenk1988} noted that halos are preferentially more prolate than oblate. Other authors \citep[e.g.][]{Bailin2005b} have found that the shape of halos varies with radius, as does the alignment of the minor axis to the disc of the galaxy.  

When defining the properties of dark matter halos a characteristic edge is required to define various parameters, particularly its mass, spin, and binding energy, etc. This can be done by defining the halo using a Friends-of-Friends algorithm \citep{Frenk1988,Huchra1982}, in which the linking length can be chosen to reliably identify matter inside an isodensity contour for appropriate parameters\footnote{A friends-of-friends linking length effectivly defines an isodensity contour given by $\Delta_{FoF}\approx 0.48L $, where $\Delta_{FoF}$ is the overdensity and $L$ is the linking length \citep{Lacey1994}. There is a detailed discussion of this in \citet{Warren2006} and in \citet{Lukic2009}.} \citep[for example][]{Frenk1988,Jenkins2001,Gott2007}. Alternatively, haloes can be defined by growing a spherical region until the mean density  exceeds a certain overdensity \citep{Press1974}. For a detailed discussion of the possible definitions of the edge of a halo see Sec. 1.2, Sec. 2.5, and Sec. 3.2 of \citet{Knebe2013} and in \citet{Lukic2009}.

As a matter of convenience, haloes are often characterised using parameters based on their virial properties. These quantities  assume spherical symmetry and a relaxed state, but provide a useful set of parameters which can be used to define the halo. 

The virial overdensity, i.e. the ratio between the mean halo density and the mean matter density of the Universe, evolves with time. A fitting function for the evolution of the virial overdensity  was found by \citet{Bryan1998}. This evolution gives a value for the virial overdensity of approximately 370 times the background density of the Universe for z=0, and a value of 222 at z=1, falling to 178 at very high redshift. The radius corresponding to the virialized region is set by the amount, and distribution, of matter within the dark matter halo, and the properties of the chosen cosmology. Other values for the `virial radius' include R$_{200}$, which is at 200 times the critical {\it or} background density of the universe, depending on the author \citep{Zemp2011}.

When the shape of dark matter halos is considered in the literature the shape profile is usually terminated at the virial radius \citep[whatever the exact definition,][]{Allgood2006}. There is on-going discussion, however, on  whether the virial radius is particularly useful in terms of the distribution of matter in the halo \citep{Kravtsov2012}.

 \citet{Prada2006} found that dark matter haloes with masses similar to the Milky Way can exhibit virialized characteristics up to 2-3 times the virial radius. Further, the evolution of the mass and virial radius of dark matter haloes of mass $\le$M$_*$ is found to be driven, at least partially, by pseudo-evolution in many cases \citep{Diemer2013, Diemand2007, Cuesta2008}. 

Based on the lack of evolution of the radial density profile, various authors \citep[e.g.][etc.]{Cuesta2008,Diemand2007,Diemer2013,Prada2006} propose that the virial radius is not a particularly special distance scale when measuring halo properties. Because the density profile extends beyond the virial radius without a strong feature at R$_{vir}$, in many cases, it will be useful to explore how the second moment tensor also changes with radius and redshift. We will probe the interactions of the outskirts of halos with the large-scale structure in what is expected (from virial theorem) to be the region of transition between the halo and the environment.

An alternative characteristic radius for a dark matter halo is the splashback radius \citep[e.g.][]{More2015}. This provides yet another definition of the edge of a dark matter halo, one which depends on the recent accretion history. In the spherical collapse model, the splashback radius is defined as the radius that separates  newly infalling material from accreted material reaching the apocentre of its first orbit \citep{More2015}, and is a natural boundary between the halo and its local environment. The splashback radius manifests as a steepening of the gradient in the halo density profile \citep[e.g.][etc.]{Diemer2014,More2015}, and varies in relation to the virial radius according to the accretion rate \citep[see Eqn. 5 in][]{More2015} and other properties.

This picture is further complicated by the fact that there is a marked difference between the spherical shape of the halo imposed by  truncating the density field at a given radius, and the more irregular distribution traced out by isodensity contours identified using friends-of-friends (FoF) methods\footnote{A discussion of the different halo finders available can be found in \citet{Knebe2013}}. This is because dark matter haloes are not spherical, but triaxial \citep{Jing2002}. \citet{Bailin2005b}, for example, find that dark matter haloes become more flattened and triaxial with radius (where $R>0.6R_{vir}$), as the haloes have not had time to fully virialize in the outer regions. The inner halo becomes more highly spherical when baryons are included in the simulations \citep[e.g.][]{Tissera1998}. Thus, we expect haloes to be increasingly triaxial with radius.

In this paper we identify features in the shape profile of galaxies in the MUGS \citep{Stinson2010} and Illustris \citep{Vogelsberger2014, Nelson2015} cosmological simulations, which persists for several Gyr. We relate the features to the interface of the filamentary large-scale structure and the spherical halo environment. 

We discuss the simulations used in this paper in section \ref{sim}. We present the results of our halo shape analysis beyond the virial radius in section \ref{results}, using the average properties of large numbers of halos taken from the Illustris simulation, and the evolution of features in individual galaxies using the higher-resolution MUGS. We discuss the MUGS simulated galaxy sample in terms of a single example halo, followed by a similar analysis of our entire halo sample (16 galaxies). Finally, we discuss our interpretation of these results and present our conclusions in section \ref{discuss}.

\section{Simulations}\label{sim}
We analyse the dark matter haloes of the 16 MUGS galaxies \citep{Stinson2010, Nickerson2013} in detail, and a larger sample of galaxies from the Illustris simulation \citep{Vogelsberger2014,Nelson2015}. The samples are  discussed in detail below, and are summarised in Table \ref{Tab:samples}.

\begin{table*}
\centering
\begin{tabular}{cccccc}
\hline
Simulation & \# halos & softening & Mass$_{dm}$ & M$_{vir}^{max}$ & M$_{vir}^{min}$  \\
 & & kpc & M$_\odot$ & h$^{-1}$M$_\odot$ & h$^{-1}$M$_\odot$ \\
\hline
MUGS (z=0) & 16 & 0.31 & 1.1$\times$10$^{6}$ & 1.6$\times$10$^{12}$ & 3.2$\times$10$^{11}$ \\
Illustris-1-Dark (z=0) & 4878 & 1.44 & 7.5$\times 10^{6}$ & 2.9$\times$10$^{14}$ & 8.5$\times$10$^{10}$\\
Illustris-1-Dark (z=1) & 300 & 1.44 & 7.5$\times 10^{6}$ & 7.8$\times$10$^{13}$ & 1.2$\times$10$^{12}$\\
Illustris-2-Dark (z=0) & 1384 & 2.88 & 6$\times 10^{7}$ & 2.9$\times$10$^{14}$ & 4.1$\times$10$^{11}$\\
Illustris-3-Dark (z=0) & 499 & 5.26 & 4.8$\times 10^{8}$ & 2.9$\times$10$^{14}$ & 1.3$\times$10$^{12}$\\

\hline

\end{tabular}
\caption{Halo samples from Illustris and MUGS used in this paper}
\label{Tab:samples}
\end{table*}

\subsection{MUGS \& MaGICC}
The MUGS galaxies have masses similar to the Milky Way, ranging from 5$\times 10^{11}$ to 2$\times 10^{12}$M$_\odot$. The simulations were carried out using the smooth particle hydrodynamics code, GASOLINE \citep{Wadsley2004} using a cosmological `zoom' simulation which includes gas and stars along with the dark matter. The dark matter, which is of primary interest in this paper, is evolved using the same method as implemented in PKDGrav \citep{Stadel2001}, which uses a kD-Tree to calculate gravity using fourth order multipole moments. We also include in our sample a galaxy from the MaGICC simulations \citep{Stinson2013}, and a MUGS simulation excluding all baryonic physics, i.e. dark matter only.

The MUGS and MaGICC simulations utilise the WMAP Year-3 $\Lambda$CDM cosmology \citep{Spergel2007}, where ($\sigma_8$, $\Omega_m$,  $\Omega_\Lambda$, $\Omega_b$, $h$)=(0.79, 0.24, 0.76, 0.04, 0.73). 
The galaxies were selected at random, but required to be of mass 
5$\times$10$^{11}<$M$_{vir}<$2$\times$10$^{12}$M$_\odot$, and not within 2.7 Mpc of a structure of mass greater than 5$\times$10$^{11}$M$_{\odot}$. 

The zoom technique then takes these randomly selected galaxies and adds higher-resolution dark matter particles in the region of interest, being careful to avoid contamination of the high resolution region with more massive particles from farther out. The high resolution region also includes baryons in the form of gas particles, which can form stars. The mass resolution of the dark matter, gas and stars is 1.1$\times$10$^{6}$M$_\odot$, 2.2$\times$10$^{5}$M$_\odot$ and 6.3$\times$10$^{4}$M$_\odot$, and the spatial resolution is 310 pc in the highest resolution region. 

The baryonic physics of MUGS uses low-temperature metal cooling \citep{Shen2010} using CLOUDY \citep{Ferland1998}; a Kennicutt-Schmidt Law \citep{Kennicutt1998} linking star formation and gas density; stellar feedback in the form of the blastwave model of \citet{Stinson2006}, which dumps energy back into the interstellar medium; and the UV background radiation is from \citet{Haardt2012}. MaGICC uses the same basic physics, but includes early radiative feedback from newly formed stars \citep{Stinson2013}. The MaGICC galaxies use the same initial conditions as the MUGS galaxies, the only difference being the implemented baryonic physics.

A complete description of MUGS and MaGICC can be found in \citet{Stinson2010,Nickerson2013} and \citet{Stinson2013}. 

In order to identify dark matter haloes and subhaloes we made use of the Amiga Halo Finder\footnote{http://popia.ft.uam.es/AHF/Download.html, \citet{Knollmann2011}} (AHF) from \citet{Knollmann2009} and \citet{Gill2004}. AHF lays an adaptive mesh grid over the density distribution of the simulation, and identifies potential (sub)halo centres as density peaks. An unbinding procedure is used to remove particles that are not bound to the density peak.The resulting halo is used to calculate halo parameters such as the virial mass and radius. Subsequently, however, we add those unbound particles back into the halo for better comparisons with other samples. The resulting haloes are considered to be composed of particles within a spherical overdense region of radius $r$, calculated from the mean overdensity of matter inside the halo. We use the $\Delta_{vir}$ value defined by the cosmology, approximated using the fitting function of \citet{Bryan1998}, rather than $\Delta_{200c}$ (which is an overdensity of 200 times the critical density of the universe), which is sometimes used in the literature \citep[e.g.][]{Navarro1996,Navarro1997}.

\subsection{Illustris}
Unlike MUGS, Illustris simulates an entire cosmological volume  106.5 Mpc on the side at fixed gravitational softening length and dark matter mass resolution. There are dark matter only and hydrodynamical versions of the same initial conditions. We make use of the $\sim$4000 most massive halos from the Illustris-1-Dark\footnote{`Dark' indicates the dark matter only Illustris runs} simulation, and complement this with the $\sim$1300 most massive halos drawn from the Illustris-2 dark matter only simulation (Illustris-2-Dark), as well as $\sim$500 halos from Illustris-3-Dark. The Illustris-1-Dark simulation has a mass resolution and gravitational softening length of $7.5 \times 10^{6}$M$_\odot$ and 1.44 kpc respectively, while the Illustris-2-Dark simulation has a mass resolution and gravitational softening length of 6$\times 10^7$M$_\odot$ and 2.84 kpc respectively etc. There are approximately 8 (64) times fewer particles in a halo of equivalent mass in the Illustris-1-Dark (Illustris 2-Dark) simulation compared to MUGS, but still a significant number within the virial radius of Milky Way sized halos. We also compare a halo from our Illustris-2-Dark sample with the same halo in the Illustris-2-Hydro simulation to explore the influence of baryons. To explore how halos evolve we additionally use a snapshot from the Illustris-1-Dark simulation at z=1, so that we can compare those results with the present day.

The cosmology used in each Illustris run is taken from WMAP Year-9 \citep{Hinshaw2013}  ($\sigma_8$, $\Omega_m$,  $\Omega_\Lambda$, $\Omega_b$, $h$)=(0.809, 0.2726, 0.7274, 0.0456, 0.704). 

The advantage of this simulation is that it does not use the zoom method utilized in MUGS. This means that there can be no artefacts due to the nested initial conditions. In zoomed simulations one could imagine that a feature might appear at the interface between more refined particles and less refined particles. In Illustris this is not an issue because all dark matter particles have the same mass and spatial resolution. 

The simulation is carried out using the AREPO moving mesh cosmological simulation code \citep{Springel2010}, and a complete description of the simulation can be found in \citet{Vogelsberger2014} and \citet{Nelson2015}. We make use of the dark matter only simulation because the baryons are not expected to strongly influence the halo shape profile in the outer regions, where the potential is dominated by the dark matter.

Halos are identified using the SUBFIND halo finder \citep{Springel2001} which utilizes a friends-of-friends algorithm to identify groups based on local density and then identifies bound substructures. 

We extract halos from Illustris, selecting the most massive subhalo from a Friends-of-Friends  group, as well as the inner fuzz (dark matter particles found by FoF but not bound to the halo) and the outer fuzz (the extra-halo dark matter in between the different halos)\footnote{ The description of particles that are part of the FoF group, but not bound to a SUBFIND subhalo as `fuzz' was used in \citet{Springel2001}. The definition of inner and outer fuzz is given in http://www.illustris-project.org/data/docs/specifications/}.  

The outer fuzz consists of dark matter particles that are not bound to any (sub)halo. We include this component in our sample out to 2.5 R$_{vir}$ for each halo. We deliberately exclude particles that are part of other friends-of-friends groups because these dense objects introduce additional peaks into our measurements, and can hide the signal of the underlying density field. Thus, the region of interest around a given halo includes: (1) the particles from the most massive subhalo for a given object, excluding other subhalos of the host FoF group (2) unbound particles in the host FoF group and (3) dark matter particles within 2.5 R$_{vir}$ that are not part of another object. For a discussion of the effect of including substructures see the Appendix \ref{App:MugIl}.

In order to calculate the late-time mass accretion history we make use of the Sublink merger tree catalogue \citep{Rodriguez-Gomez2015}, and extract the (sub)halo masses for each (sub)halo in the sample at z=0 and z=0.5. This is in line with the redshift range used by \citet{More2015} for their work on the splashback radius.

We normalize the sphericity profile using the virial radius provided by the Illustris data set, in order to match with the virial radius calculated by AHF for MUGS.  

While AHF truncates the dark matter halo at a given radius consistent with either the virial radius (for halos) or the saddle-point of the density field (for subhalos), SUBFIND traces out the density contour of the underlying density distribution, producing an irregular shape. SUBFIND uses FoF to link close dark matter particles together into a halo, before unbinding and identifying subhalos. However, a given linking length corresponds to an approximate isodensity contour \citep{Frenk1988}. This means that AHF halos consist of those particles within a distance R$_{vir/saddle}$ of the halo centre that are bound to it, while SUBFIND halos are made up of bound particles within an isodensity contour. The difference in halo properties recovered using the two methods is not large, see \citet{Knebe2011}. However, we reiterate that unbound particles are included in our halos, so long as they are not part of another object.

\section{Results}\label{results}

We first calculated the shape profiles of the dark matter haloes in our various samples.

We subdivided the haloes in our samples into spherical shells, and applied the frequently-used second moment tensor approach\footnote{This is often, erroneously, referred to as the `inertial tensor'.}. We calculated the second moment tensor,  I, of particles in a series of concentric radial bins, where

\begin{equation}
\textbf{I} = \sum^n_{i=1} m_i \textbf{x}_i \textbf{x}^T_i,
\end{equation}

\noindent
where $\textbf{x}_i$ is the position vector of particle $i$ from the centre of the halo.
The eigenvalues of {\bf I} are $I_{xx}$, $I_{yy}$ and $I_{zz}$, using the notation of \citet{Springel2004}, and the corresponding axial ratios for the shape ellipsoid are a, b, c = $\sqrt{I_{xx}}$, $\sqrt{I_{yy}}$, $\sqrt{I_{zz}}$. We also define that a $>$ b $>$ c henceforth. 

We do not use the commonly implemented iterative method, where the particle selection is deformed iteratively \citep[e.g.][]{Zemp2011}, because the matter distribution outside the halo is often strongly non-ellipsoidal. This means that the iterative approach is not appropriate because it will not converge effectively.

Using these axial ratios we further define the sphericity, $S$, where

\begin{equation}
S = \frac{c}{a},
\end{equation}
\noindent
which is the ratio of the major ($a$) and minor axes ($c$) of the shape ellipsoid, as a measure of how spherical the dark matter halo is at a given radius. 

We also measure the triaxiality parameter, $T$, first defined by \citet{Franx1991},

\begin{equation}
T = \frac{1-b^2/a^2}{1-c^2/a^2},
\end{equation}
\noindent
which measures the degree to which $a$, $b$ and $c$ are different. A halo is considered oblate where $0<T<1/3$, triaxial where $1/3<T<2/3$ and prolate where $T>2/3$. 

We measure the shape in spherical shells extending out to 2.5 R$_{vir}$, but measure the density out to 5 R$_{vir}$. This was done so ensure we include the splashback radius in the density profile \citep[e.g.][]{More2015}.

The splashback radius is defined in terms of the minimum of the density gradient, specifically,

\begin{equation}
y = \frac{d\log(\rho)}{d\log(r)}
\label{Eqn:More}
\end{equation}

\noindent
where $\rho$ is the density in a given spherical shell and $r$ is the radius.

\citet{More2015} identify a correspondence between the splashback radius relative to R$_{200b}$ (the radius marking out an overdensity of 200 above the background (matter) density of the Universe), that depends on the infall rate. This is parametrized in terms of $\Gamma$, where,

\begin{equation}
\Gamma = \Delta \log(M_{vir}) / \Delta \log(a_{exp}),
\end{equation}
where $M_{vir}$ is the halo virial mass and $a_{exp}$ is the expansion factor. Masses are taken from merger trees at z=0 and z=0.5 by following the halo main branch. 

Although \citet{Diemer2014} and \citet{More2015} define the minimum of the density gradient profile to be the splashback radius, we will point out that although the radius itself is hard to define in individual halos (with considerable noise due to temporary fluctuations in the halo, such as subhalos or transient streams etc.) there is a long term splashback {\it region} which evolves smoothly with time (Section \ref{MUGS}).

\subsection{Average Shape Profile for Halos in Illustris}

\begin{figure}
    \centering
     \includegraphics[scale=.4]{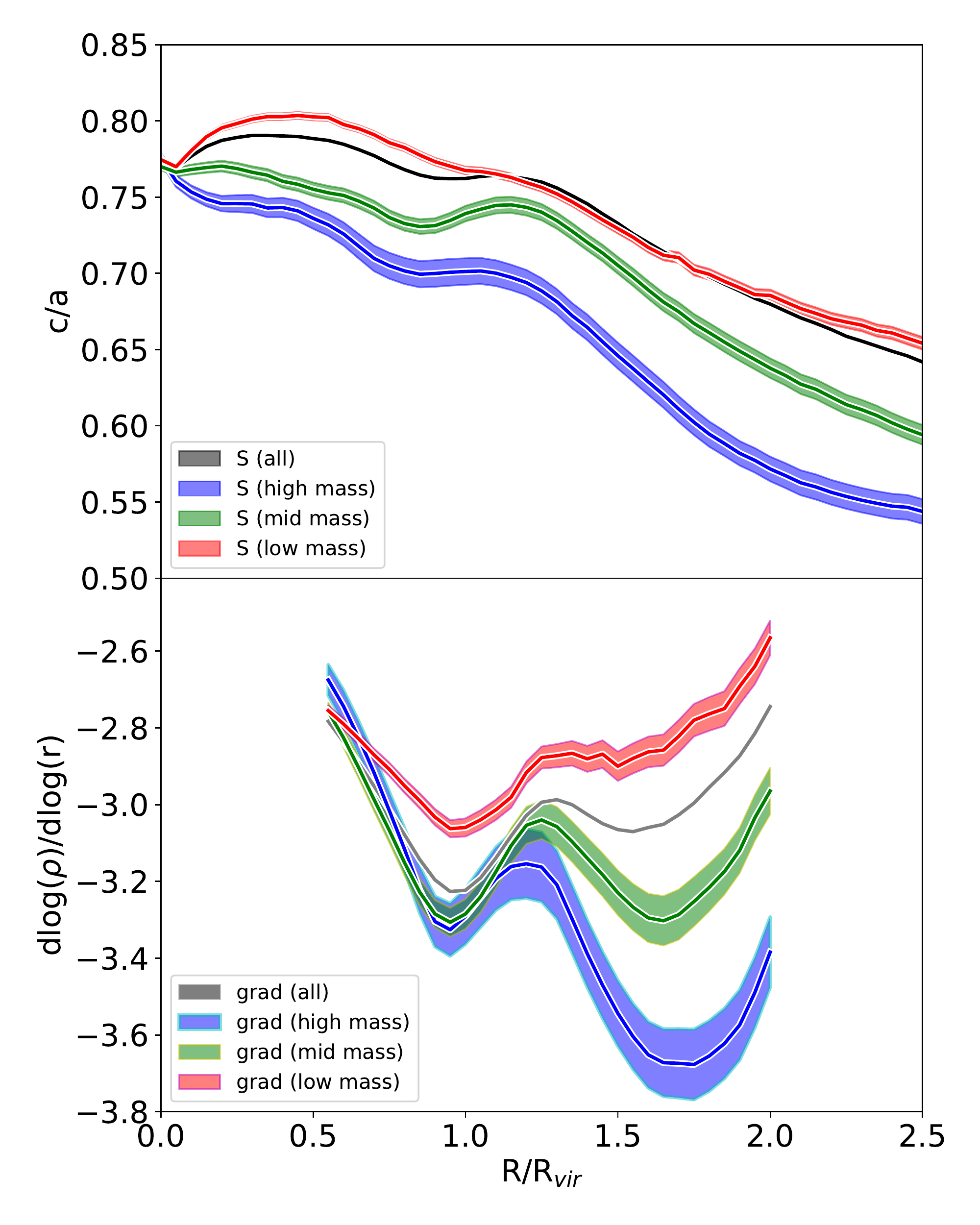} 
     \caption{The mean sphericity (top panel) and density gradient (bottom panel) profiles of dark matter halos extracted from the Illustris-1-Dark sample. The different lines show different subsamples, while the shaded regions are the corresponding standard error ($\sigma/\sqrt{N}$). In the legend 'S' shows the lines for sphericity, while grad indicates the density gradient. The gradient profiles are only shown between 0.5 R$_{vir}$ and 2 R$_{vir}$. There is a correspondence between the dip and recovery in the sphericity and the dip and recovery in the density gradient profile close to R$_{vir}$.}
 \label{Fig:Averages}    
\end{figure}

\subsubsection{Sphericity properties}
In order to identify the `global' shape of the sphericity profiles in Illustris it is useful to stack the profiles of halos. 
Sphericity is used rather than triaxiality because the triaxiality signal is lost for lower mass halos (below $10^{13}$M$_\odot$), even in the highest resolution sample. The signal can, however, still be seen in sphericity.

 Using the halos from our Illustris-1-Dark sample we plotted the  mean halo shape profile. We normalize the profiles by the virial radius, and take the mean of the sphericity profiles of all halos in each normalized radial bin. 
 
 Figure \ref{Fig:Averages} shows the sphericity profile of the entire sample from Illustris-1-Dark.  The sample is also subdivided into different mass bins. The highest mass bin contains the mean sphericity profile for halos with mass $> 10^{13}$M$_\odot$, the intermediate mass bin contains halos with mass $10^{12}$M$_\odot<$M$< 10^{12.5}$M$_\odot$ and the lowest mass bin contains objects with masses of $10^{11}$M$_\odot<$M$< 10^{11.5}$M$_\odot$.

 There are two distinct regions in the sphericity profile. Specifically, (1) the inner region; which has a low sphericity gradient (inside 0.6 R$_{vir}$) and terminates in a `dip and bump' around the virial radius, the exact location of which depends on halo mass, and (2) an outer region; where the sphericity drops away more rapidly outside 1.3 R$_{vir}$. The dip and recovery for the whole sample occurs at radii of approximately 0.92 and 1.1 R$_{vir}$ respectively. The `bump' feature in the sphericity around the virial radius is evident in both of the more massive mass cuts, as well as in the whole sample, within the standard error on the mean \citep[also see][]{Suto2016}.

This demonstrates that there is no single object or halo mass subset which generates the features, but that it is a general result. It is also evident that more massive halos are less spherical (with sphericities of 0.7, 0.74, 0.77 for the most to least massive samples). The whole sample has a mean sphericity of 0.74), as found by previous authors \citep[e.g.][]{Jeeson2011}. The lowest mass bin has too few particles per bin to accurately identify the dip above the noise (around 1000 particles per bin at R$_{vir}$), because of increased bin-to-bin scatter in the recovered shape (discussed in detail in the Appendix \ref{Append_res}).  Although the dip feature is not identifiable at this resolution, the overall profile is well reproduced. This demonstrates that a sufficient number of particles is required to identify the dip feature, and this must be considered when comparing this result to other works. The number of particles needed to measure the halo shape is larger than required to identify radially averaged quantities.

The  mean halo shape in the Illustris sample experiences a `dip and rise' feature in sphericity at the virial radius. It is strongest in the intermediate mass bin with a change in sphericity of 0.015 between the bottom of the dip and the top of the rise. This means that just inside the virial radius the halo becomes less spherical, then more spherical again at or close to the virial radius, before becoming increasingly non-spherical in the local environment, as expected. In the other mass bins the feature manifests as  a flattening of the shape profile for $\sim$0.2 R$_{vir}$.

\begin{figure}
    \centering
     \includegraphics[scale=.4]{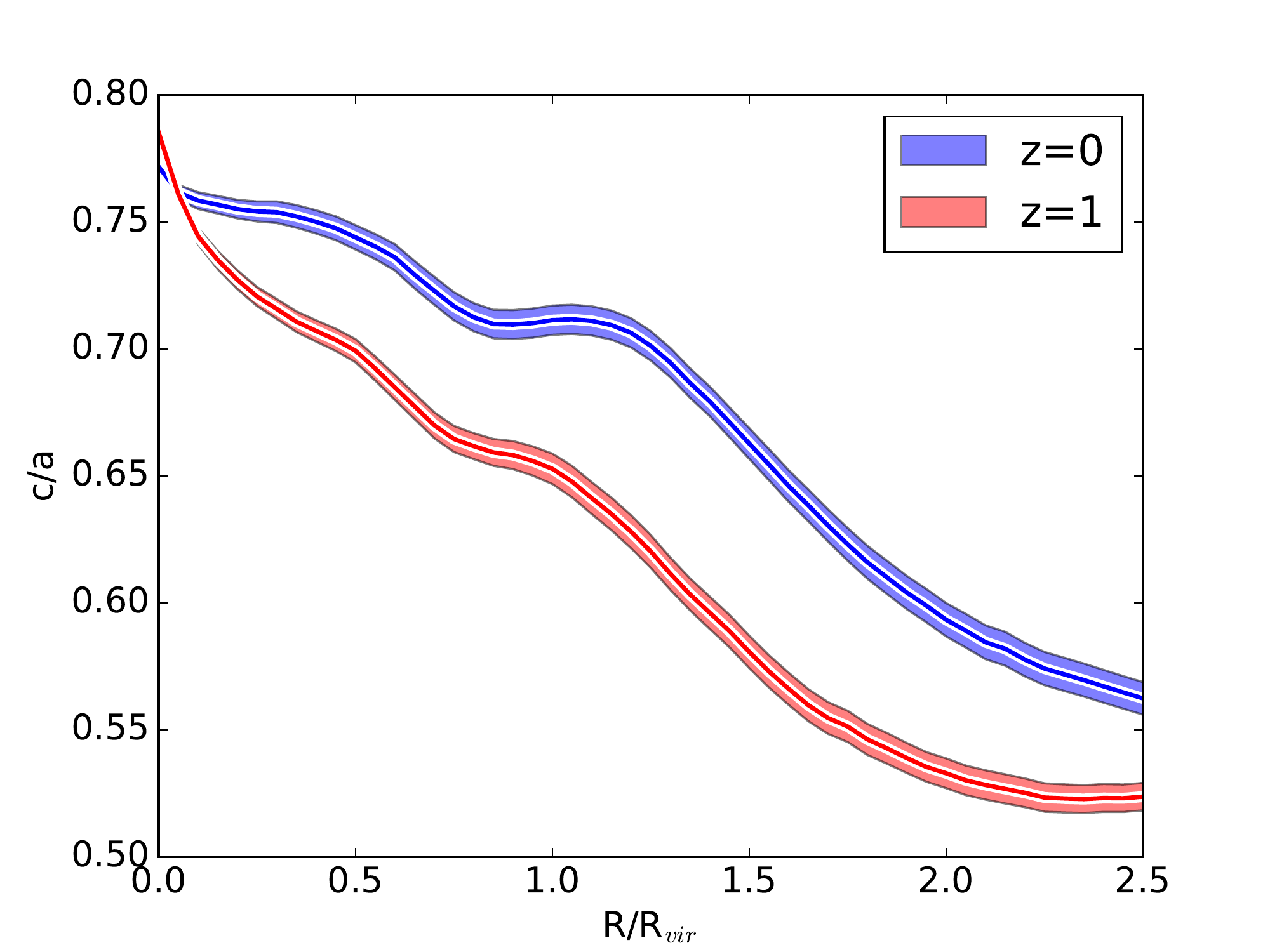} 
     \caption{The sphericity profile of the Illustris-1-Dark `average halo' at z=0 (blue) and  z=1 (red) for the 300 most massive objects. The dip feature is present in each snapshot.}
     \label{Fig:TimeevolIllusrtis}    
\end{figure}

 Although we have found a robust structure in the `average' dark matter halo in Illustris, in order to examine the persistence of the structure we also explore the time evolution between z=0 and z=1, using snapshots at both of these epochs. Various authors \citep[e.g.][]{Allgood2006,Cuartas2011} have observed that halo shapes vary with redshift, becoming more spherical at later times, and we see the same in Fig. \ref{Fig:TimeevolIllusrtis}. At z=1, the sphericity of the halo ensemble at the virial radius is 0.06 lower than at z=0. 
 
Figure \ref{Fig:TimeevolIllusrtis} shows the shape profile for the 300 most massive halos in Illustris-1-Dark at z=0 and z=1. The mean profile at z=1 shows the same feature as at z=0, even if the signal is weaker. This illustrates that the dip in the shape profile is an important characteristic of halos for a considerable fraction of the age of the Universe. By comparing these two redshifts we see that the shape feature remains close to the virial radius even as the universe evolves, and that it is {\it persistent}. Although the feature is present at early times it is reduced in size, implying that evolution is taking place. The virial radius grows with time  partly because the expansion of the Universe lowers the background/critical density  \citep[called pseudo-evolution,][]{Diemer2013}. The shape feature appears to keep pace with this growth in radius (see Fig. \ref{Fig:g15784zrevolsph} and \ref{Fig:sphfeaturemugs}). 
The flattened region in the shape profile declines from a width of 0.2 R$_{vir}$ to almost $\sim$0.1 R$_{vir}$ with increasing redshift. It has also moved inwards, ranging from 0.75-1 R$_{vir}$.  However, the shape profile has a steeper gradient (-0.1) at z=1 compared to z=0 (-0.04), which may act to hide the feature. 
  
While some of this may be due to the increased number of particles per bin as the halos grow (Appendix \ref{Append_res}), the average number of particles in the R=R$_{vir}$ bin only changes by 30\%. Furthermore, the effect is also seen in samples chosen to have the same number of particles at z=0 and z=1, indicating that the feature is not purely due to resolution.   

\subsubsection{Sphericity and splashback}
\citet{Diemer2014} and \citet{More2015} examine the gradient of the density profile of dark matter halos in simulations, and explore the properties and evolution of the `splashback radius' using the gradient profile defined in Eqn. \ref{Eqn:More}. They find a change of gradient at a given radius which depends on the rate of infall into the halo. 

For the Illustris-1-Dark sample  we compare the location of the average plateau/dip feature in the shape profile with the trough in $d\log(\rho)/d\log(r)$ (which we will call the splashback region). We followed \citet{More2015} in that we smoothed the density profile using a  \citet{Savitzky1964} filter over 15 neighbours and a 4th order polynomial. We then took the average of the density profiles in the sample. We differ from \citet{More2015} in that we removed substructures in order to reduce bin-to-bin noise. \citet{Mansfield2016} demonstrate that substructure can strongly change the apparent density profile of halos, and make identifying the splashback radius more difficult. A discussion of the impact of this is in Appendix \ref{App:MugIl} and Fig. \ref{Fig:shapesplashSH}.

 The mean density gradient profile is shown alongside the mean sphericity profile for the Illustris-1-Dark sample in the lower panel of Fig. \ref{Fig:Averages}. In Fig. \ref{Fig:Averages} there is a clear `two dip' shape, with local minima at 1 and 1.7 R$_{vir}$, not observed in   \citet{More2015} or \citet{Diemer2014}.  This is the result of caustics coming from already accreted material undergoing its second orbit, and from infalling material undergoing splashback. The true splashback radius is the one further out \citep{Adhikari2014}. We discuss this in detail below, and in Appendices \ref{Append_slp} and \ref{App:MugIl} in particular. 
 
For the whole sample, the strongest dip is close to the virial radius (and the inner caustic, which is {\it not} the splashback radius), while in the highest mass subset it is at 1.7 times the virial radius (the true splashback radius). This is the opposite to what is observed in \citet{Adhikari2014}, who find that the outer caustic is deeper than the inner caustic.   
It may be due to our removal of substructure. An in-depth study of this is beyond the scope of this paper. The local minimum of the density gradient close to R$_{vir}$ corresponds to the dip in the sphericity profile,  indicating the seperation of virialized material from infalling material. In \citet{Adhikari2014}, the inner caustic corresponds to the second orbit of accreting material. 

In high mass halos we recover the result of \citet{More2015}, that as the accretion rate increases the splashback radius moves inwards. For all masses, and at low accretion rates, two distinct dips in the density gradient profile develop due to the development of two distinct caustics \citep[as in][]{Adhikari2014}. 

In \citet{Diemer2014} the two caustics are not observed, except for very low values of $\Gamma$,  \citep[see Figures 7, 10 and 14 of][]{Diemer2014}, while it is much stronger in our sample. Our splashback radii are calculated after the removal of substructure, while \citet{Diemer2014} did not do this. The inclusion of substructure adds additional peaks to the profiles, which can hide details such as the two caustics, although not completely (see Fig. \ref{Fig:shapesplashSH} and \citet{Mansfield2016}). However, as we are interested in the underlying shape of the parent halo density distribution we remove these objects from the sample.

{\it The beginning of the splashback region (where the density gradient begins to fall) corresponds to the final decline of the sphericity as the halo blends into the filamentary large-scale structure. This strongly suggests that the feature in the sphericity and the `splashback' radius in the density profile are related \citep[e.g.][]{Suto2016}.}

\begin{figure}
    \centering
    
     \begin{tabular}{cc}

     \includegraphics[scale=.38,trim={0.2cm 0.5cm 1.6cm 0.9cm},clip]{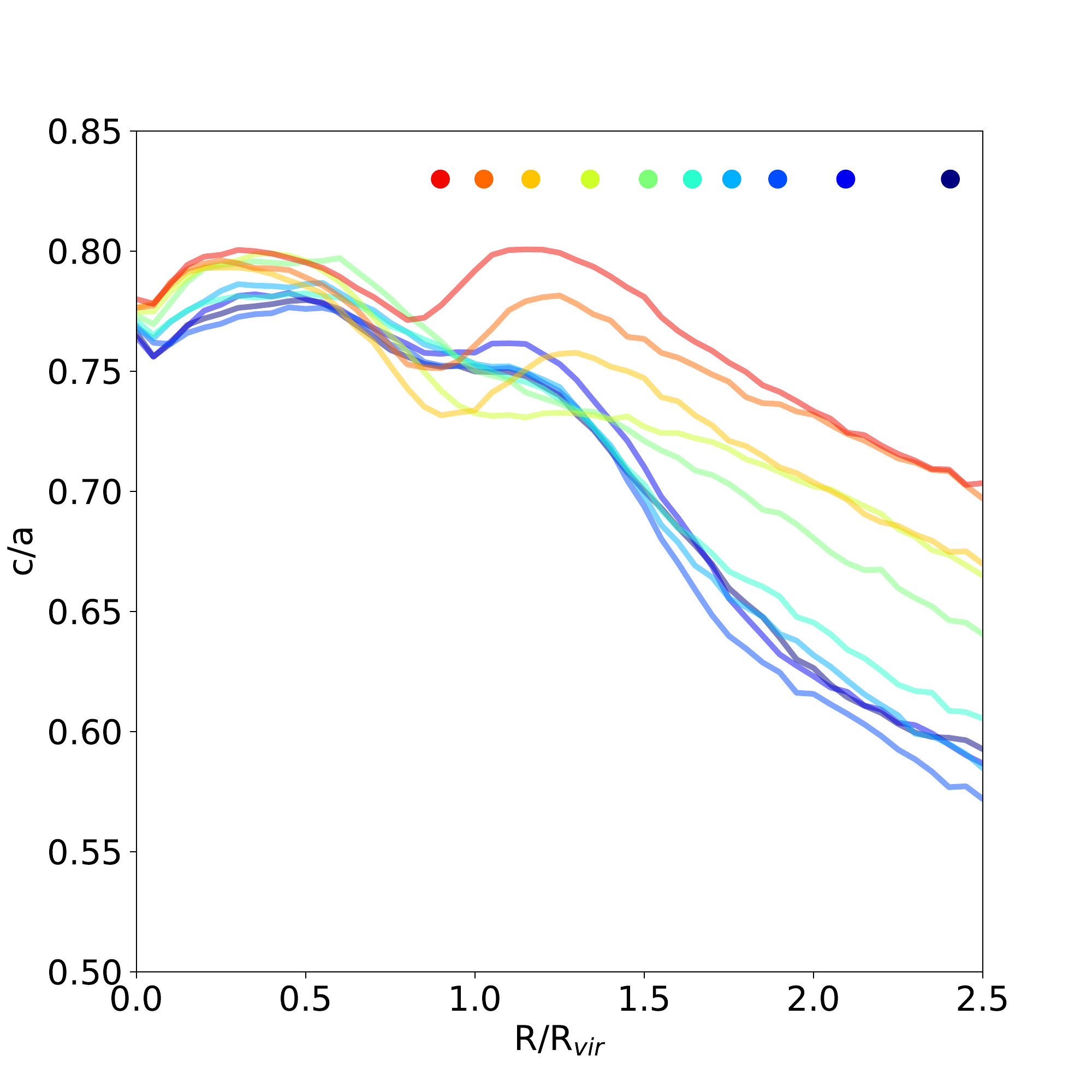} \\

     \includegraphics[scale=.38,trim={0.2cm 0.5cm 1.6cm 0.9cm},clip]{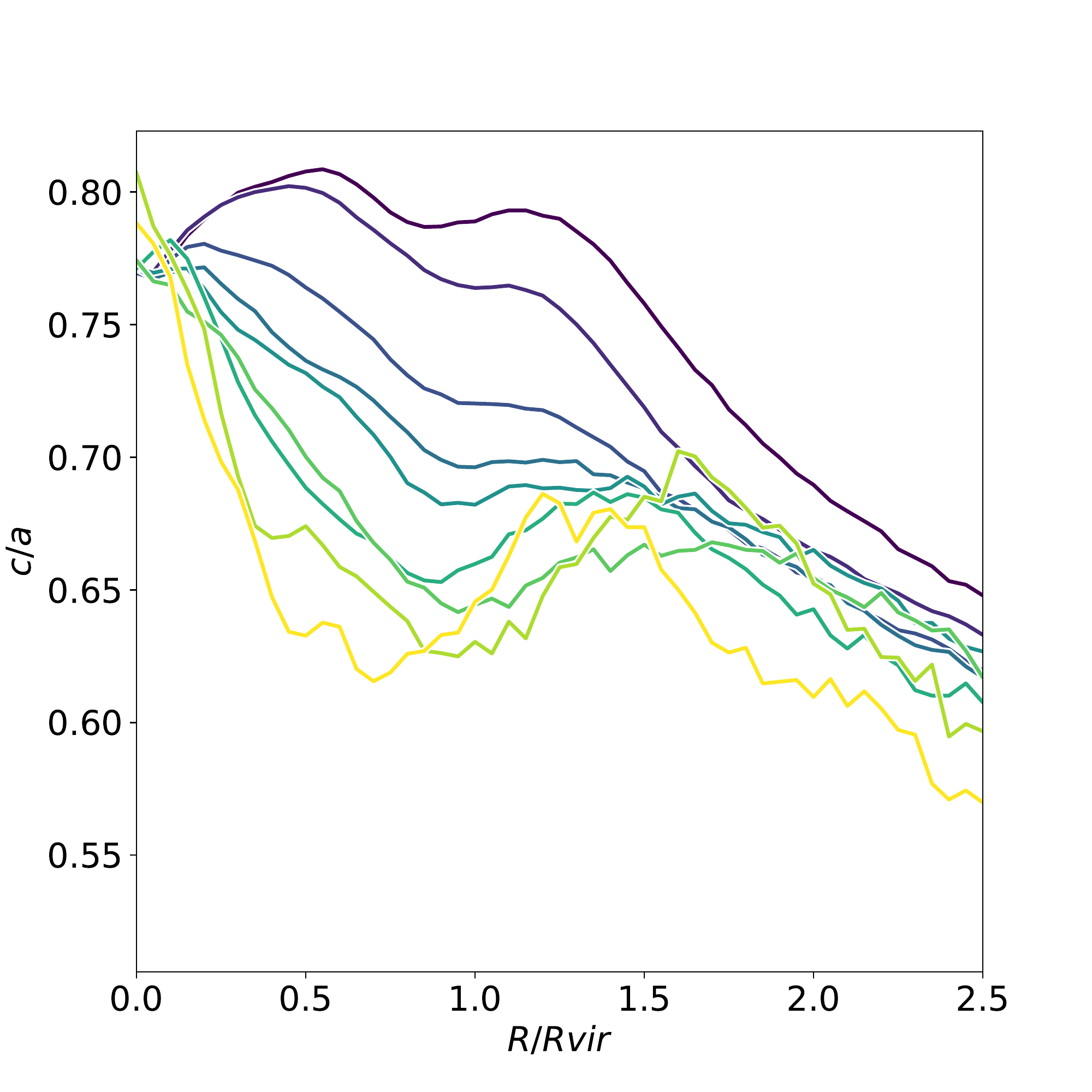} &
     \includegraphics[scale=.3,trim={.7cm 0.5cm .5cm 0.9cm},clip]{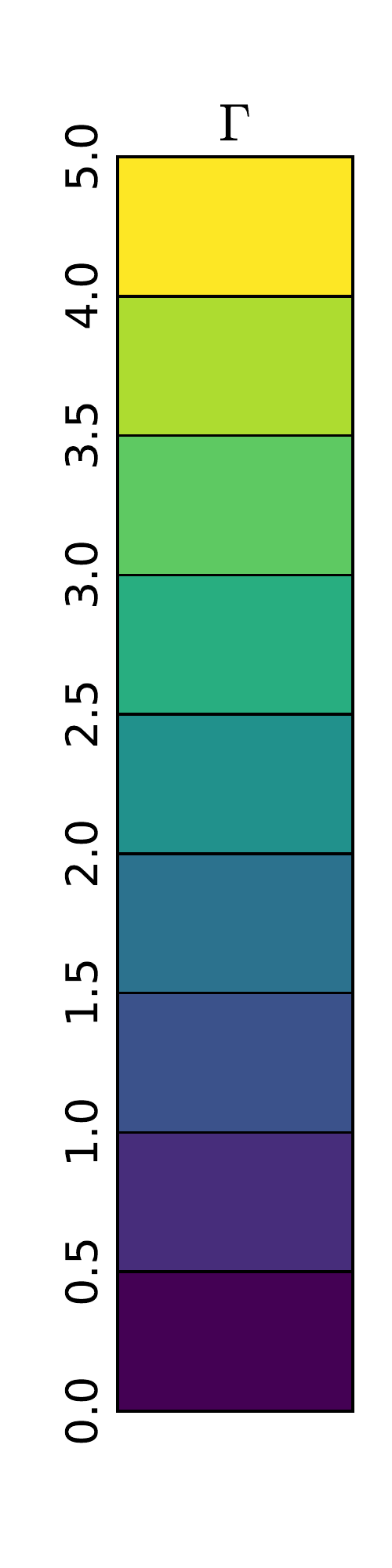}\\
        
     \end{tabular} 
     \caption{Top panel: The sphericity profiles of the Illustris-1-Dark sample binned by splashback radius. Each bin contains 300 halos ordered by the minimum of the density gradient between 0.5 and 2.5 R$_{vir}$. The mean splashback radius is given by the coloured points.  The sphericity dip is strongest for halos with a low splashback radius. Lower panel: The sphericity profiles of the Illustris-1-Dark sample subdivided by $\Gamma$ (infall rate).  The sphericity dip deepens  and moves inwards for high $\Gamma$. Halos with high $\Gamma$ are less spherical.}
     \label{Fig:Illustrisbinedsplash}    
\end{figure} 

 Where we subdivide the Illustris-1-Dark sample by $\Gamma$ we find that the trend of decreasing splashback radius with infall rate is not as clear as in previous work. However, this may be due to our sample size of just 4000 objects instead of the samples available in \citet{Diemer2014} and \citet{More2015}.

Using the Illustris-1-Dark sample we subdivided the halo sample into bins of 300 halos in order of the splashback radius, and calculated the mean sphericity profile of each subset (top panel, Fig. \ref{Fig:Illustrisbinedsplash}). Where the splashback radius is close to the virial radius the halo shape feature is strong, especially where the splashback radius and the virial radius coincide. However, the feature becomes weaker at intermediate values of R$_{splash}$/R$_{vir}$. We emphasize again that there is a large amount of variation between individual halos for both the density gradient and sphericity profiles, and this must be taken into account when comparing this result with stacked profiles such as in \citet{More2015}.

There is a correlation between the size of the dip and the infall rate, indicating a  relationship between the sphericity feature and the splashback radius (see the bottom panel of Fig. \ref{Fig:Illustrisbinedsplash}). The shape feature becomes a dip once $\Gamma$ increases above 2.5, and high infall also reduces the sphericity of halos at all radii.
\citet{Mansfield2016} find no correlation between shape at the splashback shell and $\Gamma$, but use more complex geometries than the best-fit ellipsoid and radial binning used here, i.e. their splashback shells are not at a given radius but have a lobed shape \citep[see figure 2 of][]{Mansfield2016}.

Furthermore, the shape feature moves inwards with increasing $\Gamma$, reaching 0.6 R$_{vir}$ for $\Gamma$=5. This strongly implies that the effect is dominated by the accretion rate, and that the shape feature is a result of the interplay between newly infalling and virializing material. In order to make this effect strong a considerable amount of material must be infalling.

\begin{figure*}
    \centering
     \includegraphics[scale=.7,trim={1cm 1cm 0cm 2cm},clip]{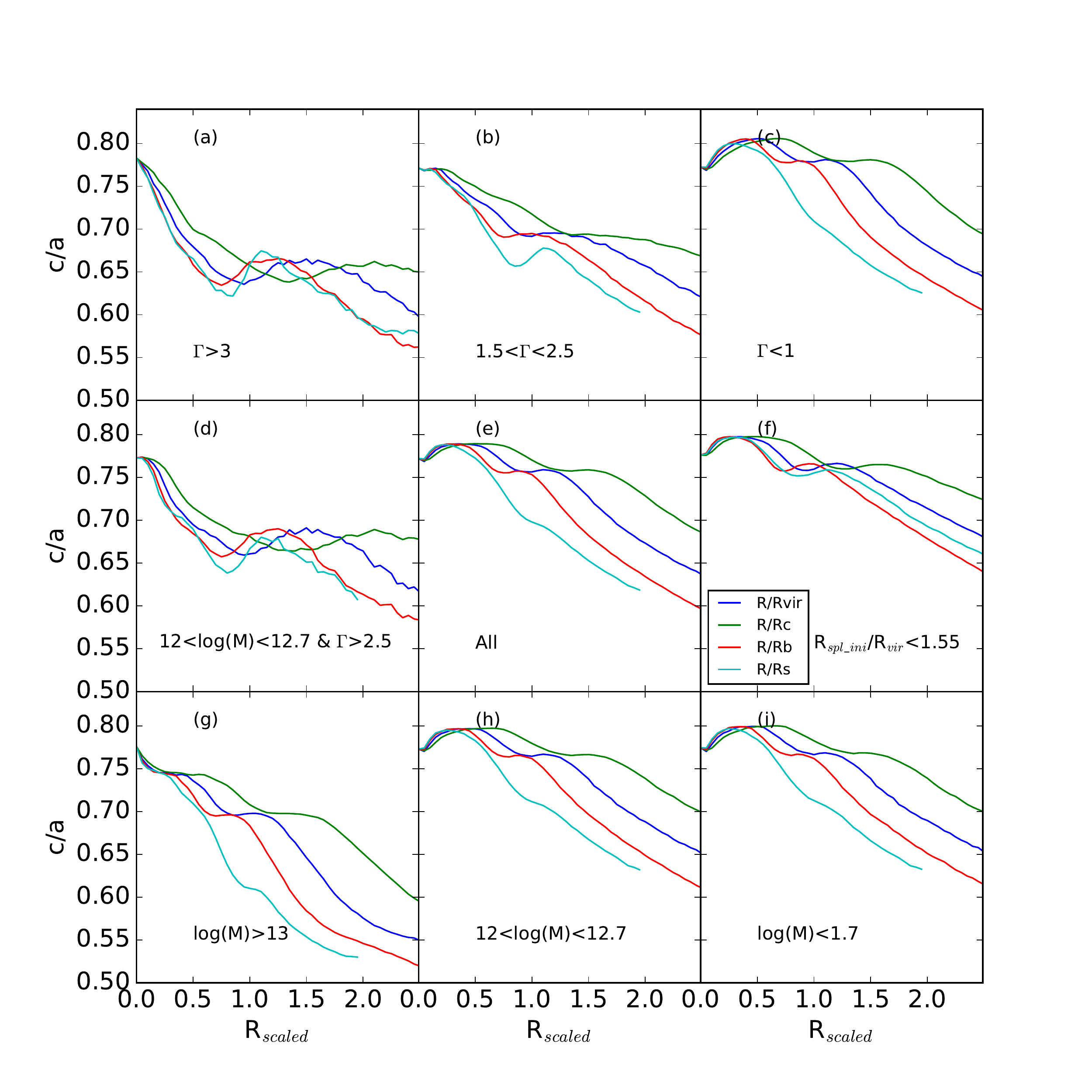}
     \caption{The shape profile of halos scaled by R$_{vir}$, R$_{200c}$, R$_{200b}$ and R$_{splash}$ and subdivided into different mass and $\Gamma$ bins. There is always a feature in the shape profile when it is scaled by a radius based on overdensity, but the size of the feature changes most strongly when the profile is scaled by the splashback radius, in different $\Gamma$ and mass bins.} 
     \label{Fig:splashcomp}    
\end{figure*}

 The sphericity trends in different mass bins for different values of $\Gamma$, and rescaled by the different `virial' radii, are shown in Fig. \ref{Fig:splashcomp}. The virial shape feature is close to 1 R$_{vir}$ and not considerably stronger in R$_{200c}$ and R$_{200b}$. Figure \ref{Fig:splashcomp} shows a strong correlation between the virial shape feature and the splashback radii at high to intermediate $\Gamma$ (panels (a) to (c)). This can be seen where the shape profile is scaled by the splashback radii (cyan line), and is also shown in Fig. \ref{Fig:Illustrisbinedsplash}.

In Fig. \ref{Fig:splashcomp}, each shape profile is scaled separately by its Rvir,
R$_{200c}$, R$_{200b}$ or R$_{splash}$. In each panel the shape of the profile is not
much different when the radius is scaled by the overdensity-based
radii. This is because there is a relatively
fixed ratio between R$_{vir}$, R$_{200c}$ and R$_{200b}$.
The shape is considerably
different when the individual profile is scaled according to the
splashback radius. This is because the profiles shown are the mean profiles
of the different halos, and so the scatter in the position of the dip
affects the resulting curve.  When we stack the individual sphericity profiles there is little change in the spread of sphericity at a given radius.  Where the shape features align, the
resulting mean profile shows a strong and narrow dip, but where the
scatter is larger the dip is shallower and broader. The significance of
the dip is more variable where the individual profiles are scaled by the
splashback radius. For example, the
shape feature can be seen in high $\Gamma$ objects when the individual
profiles are scaled by the splashback radius, indicating that the spread in sphericity at a given radius changes.

The sphericity dip is much less apparent
in the lowest $\Gamma$ bin (panel c) than in the highest or intermediate bins (panels a and b). The dip is, however, still apparent at low
$\Gamma$ when the profiles are scaled by the overdensity-based radii.
This implies that there is a correlation between the position of the dip
and the virial radius. We suggest that this is evidence of the importance of the virial radius to the shape profile. If the dip was mainly correlated with the
splashback radius we would expect the opposite behaviour, where the shape
feature would be change more in different bins when scaled by the virial radius than the splashback radius.

When we bin by mass instead of $\Gamma$ (panels (g) to (i)) the dip is weakest in low mass objects when the shape profile is scaled by the splashback radius. This is similar to the shape profile for halos with low $\Gamma$ (shown in panel (c)), which are numerous  throughout the sample.

In summary, the dip is more consistent where it is scaled by the virial radius than when the profiles are scaled by the splashback radius. This implies a correlation between the shape feature and the virial radius.

Furthermore, the sphericity of halos increases with decreasing infall rate, which suggests that high infall perturbs the halo, and means it is no longer virialised in the outer regions.

The existence of the penultimate shell in the phase space distribution (which gives rise to the inner dip of the density gradient profile) coincides with both the virial radius and the shape feature in Fig. \ref{Fig:Averages}. As evidenced in their Figure 9, \citet{Adhikari2014} note that at low $\Gamma$ the stream of material which is `splashing back' is distinctly seperated from the rest of the virialized halo material in phase space. At higher $\Gamma$ the seperation between the second to last and outer shells of the caustic can no longer be identified \citep[e.g.][]{Adhikari2014}, but the physics is unchanged.

Strongly inflowing material will disrupt the outer edge of the halo.  This has the effect of reducing its sphericity, and means that the outer regions will no longer be virialized. This also explains how the dip can be present  when the splashback radius is inside R$_{vir}$, as shown for high $\Gamma$. The high infall rate means that the material at the virial radius is perturbed and no longer virialized, so the edge of the virialized region moves inwards (the effective radius at which the halo is virialized is less than R$_{vir}$). 

The inneredge of the sphericity dip moves inwards because there is a stronger interaction between newly inflowing material and the in-situ material,  meaning that as well as moving inwards the dip in sphericity also becomes stronger. This scenario can be seen in the lower panel of Fig. \ref{Fig:Illustrisbinedsplash}, for the high $\Gamma$ (yellow) profile. The splashback and radius of virialization (which is not necessarily the formal virial radius) are intrinsically linked, even if the relationship is complicated. The outer edge also moves inwards with high $\Gamma$, because the splashback radius also moves inwards when the infall is high, as found by \citet{More2015}.

\subsection{Detailed study using MUGS}
\label{MUGS}
In this section we will explore the properties of the halo shape and density profile of a small number of high-resolution halos from MUGS, and trace them through time. We will start with a detailed examination of a MUGS halo where the sphericity dip is particularly strong, and then expand to the entire sample.

An important caveat to the following discussion is that there is a large halo-to-halo variation in the details of the shape feature and the splashback radius, Thus, more than 16 objects are required for the general trends to emerge with certainty. 

For a summary of basic galaxy properties see Table 1 of \citet{Nickerson2013}. The shape profile of g15784 is an ideal case, wherein properties of the individual halo match the average (except for a lower than expected value of $\Gamma$). In this way g15784 is a very representative halo.

\subsubsection{A view of one halo}

We discuss the MUGS sample because it is higher resolution than the Illustris-1 or 2 samples. As discussed, there is considerable object-to-object variation in the exact shape profile of different halos, ranging from negative gradients, positive gradients, peaks, plateaus and dips.  Our chosen object, g15784, shows remarkably similar splashback and shape profiles to the `average' halo from Illustris. It is not the only galaxy in the MUGS sample to show these features, just the strongest and most long-lived.

 The MUGS galaxy g15784 has a mass of 1.5$\times 10^{12} M_{\odot}$ with a very quiescent merger history. As such, we expect the halo to be fairly relaxed. The mean dark matter mass accretion rate is around 90 solar masses per year, but lower at later times (the $\Gamma$ value is 0.5, which is low for halos with a large dip, e.g. Fig. \ref{Fig:Illustrisbinedsplash}, but this is due to the large halo-to-halo variation.).

This galaxy has been discussed in other respects in a number of other papers \citep[e.g.][]{Pilkington2012, Nickerson2011, Walker2014, Calura2012, Snaith2015}, although the primary focus has been on its baryonic properties. It has a thin disc and a low bulge-to-disc ratio \citep{Stinson2010}. 

A feature is apparent close to the virial radius, characterized by a fairly sharp drop in sphericity and a subsequent recovery. This is co-incident with the feature in the average profile from Illustris, discussed previously. The dip in sphericity at 0.7-1.5$R_{vir}$ (200-490 kpc at z=0) is a persistent feature, lasting from z=1 ($\sim$6 Gyr) to the present day (13.7 Gyr). Further, the dip visibly deepens over time, starting as a brief plateau in the declining sphericity profile, and becoming noticeably deeper at low redshift  (reaching a maximum depth ranging from approximately 0.9 to 0.55 at z=0). This is, again, consistent with the mean Illustris shape profile. The transition between the dip and the  more spherical material at its outer edge coincides with the splashback radius between z=0 and z=1. However, the recovery feature is not as persistent as the dip itself, and spreads out significantly towards z=0.  This implies that something important is happening \textit{close to the virial radius and/or splashback radius of this galaxy}. The dip is not a narrow feature, but a transition region between the virial and splashback radii. The shape evolution within 0.7$R_{vir}$ does not significantly evolve over time, but exterior to 1.5$R_{vir}$ the shape becomes increasingly spherical towards the present day. This can be interpreted as a widening of the bump which occurs after the dip, a build up of spherical material outside the splashback radius, or an evolution of the large-scale structure. The shape uncertainty calculated from bootstrapping the data is small compared to the size of the feature. The number of particles in each radial bin for this galaxy does not drop below 2000 particles and is usually much higher. 

 MUGS is a zoom simulation, which means it uses dark matter particles of differing masses in concentric regions  of decreasing mass centred on the galaxy of interest. As such there is always a possibility that the region of interest will be contaminated by particles with significantly higher masses. However, we find that contamination of the region of interest with high mass particles is low. Although a few (15 within 2$\times$R$_{vir}$) particles with higher masses do lie in the outer edge of the region of interest, these high mass particles have mass only 20\% greater than the highest resolution particles, so their impact will be minimal. There are no particles from higher mass generations within the region of interest.

It is remarkable that this dip feature remains at the virial radius for over half the age of the universe, during which time the virial radius of the galaxy has increased by almost 90\% (from 177 kpc at z=1 to 330 kpc at z=0). This implies that it is not pseudo-evolution, but a feature of the galaxy, that lies at the edge of the expanding virial radius, for almost 8 Gyr, even as the virial radius moves outwards. 

\begin{figure}
  \begin{tabular}{c}
     \includegraphics[scale=.45]{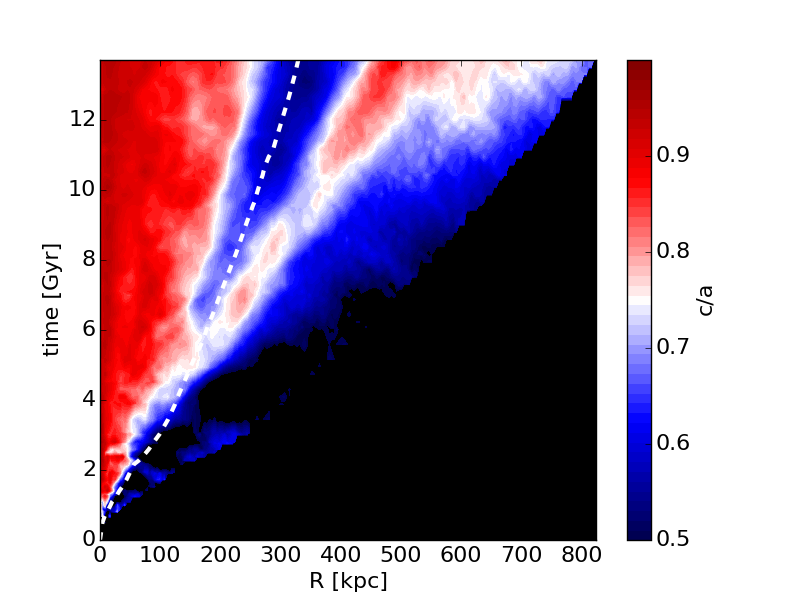}  \\
     \includegraphics[scale=.45]{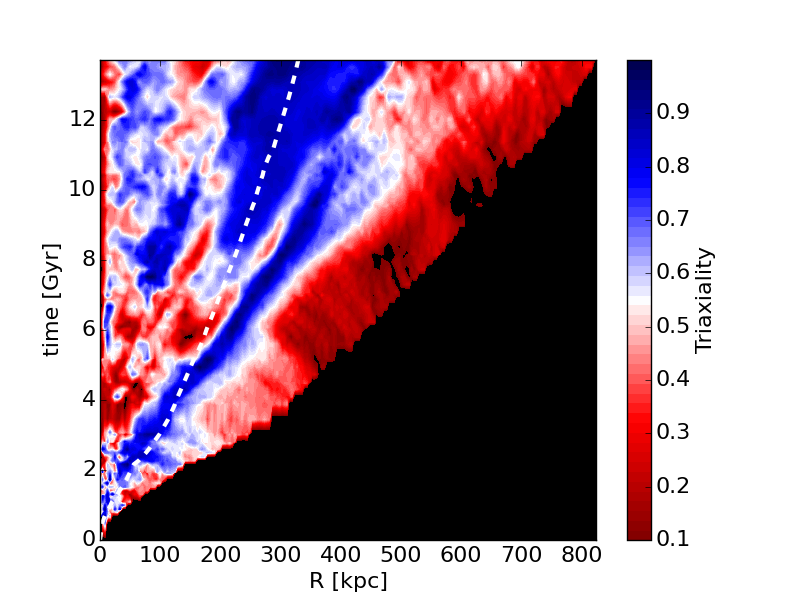}  \\
   
     % trim={<left> <lower> <right> <upper>}               
\end{tabular}
        \caption{The time evolution of halo sphericity (top) and triaxiality (bottom) as a function of physical radius for g15784,
		plotted out to $2.5~R_{vir}$. The dashed white line is the virial radius, and time increases upward
		in the plots. The virial shape feature is the blue region that
	tracks the virial radius.}

 \label{Fig:g15784zrevolsph}    
\end{figure}

Figure \ref{Fig:g15784zrevolsph} shows the change of the sphericity and triaxiality in terms of time and radius. The virial radius evolves considerably between 6 and 13.7 Gyr, and the position of the dip (in sphericity) clearly moves outwards along with it.  This drop in sphericity corresponds to an increase in the degree of `prolateness' of the halo at the virial radius, implying that one of the three principle axes is emphasised above the other two, even while the radial density distribution is unaffected. 

\begin{figure}
    \centering
     \begin{tabular}{c}
     \includegraphics[scale=.45]{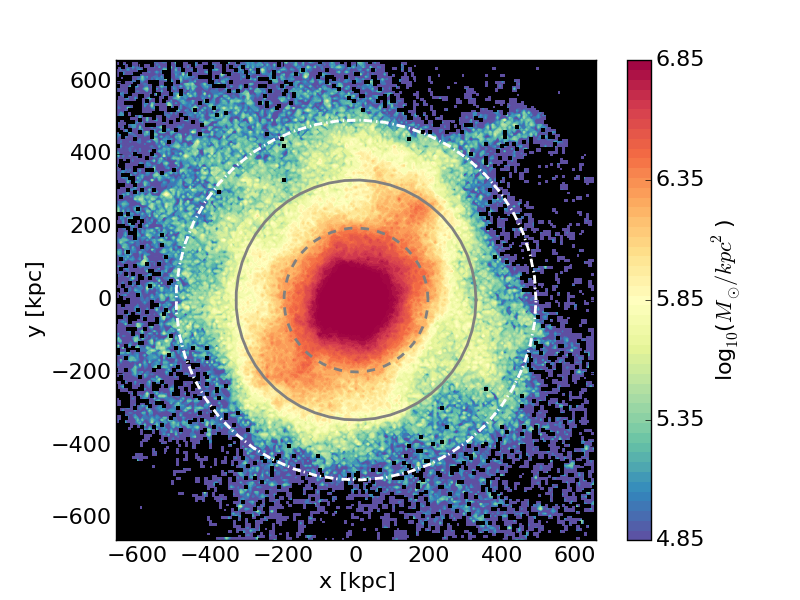} \\
     \includegraphics[scale=.45]{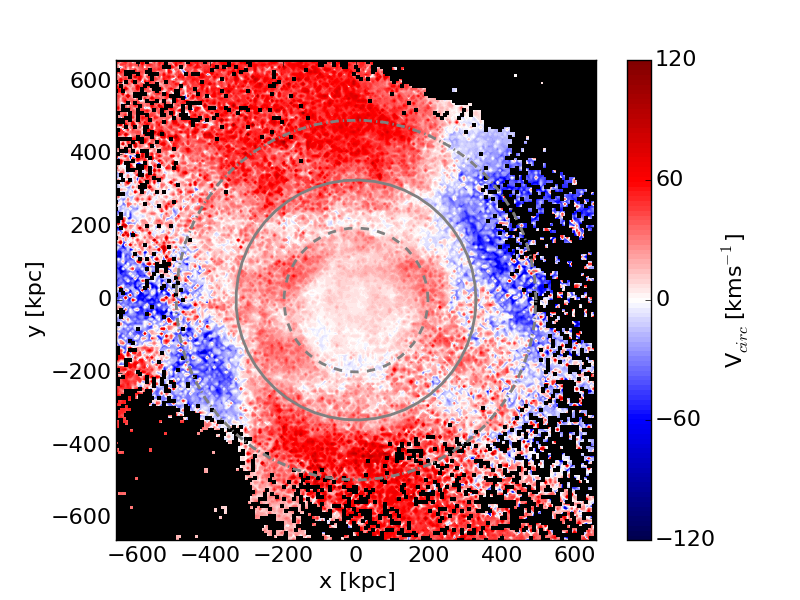}
     \end{tabular}
     \caption{Top: Projected distribution of matter around g15784, coloured by surface density at z=0 oriented so that the galaxy disc is face on. Bottom: Bulk angular flows of the dark matter, in the direction orthogonal to the angular momentum of the galaxy disc, described in terms of the circular velocity distribution (defined as $\sum_i(m_iv_i\cdot r_i)_z/\sum_i(m_ir_i)$ in each pixel). Inside the virial radius, subhaloes are discarded, while outside the virial radius of g15784  dark matter belonging to other haloes are discarded, leaving only the background. The grey circles mark 0.6, 1 and 1.5 times the virial radius respectively. These distances correspond to the beginning, centre and end of the dip feature. The halo is oriented such that the galaxy disc is aligned to the z-axis.The box is cubic, with the same cuts along the z-axis as in x and y, each axis is $\pm$2R$_{vir}$. }
 \label{Fig:g15784density}    
\end{figure}

In order to address the origin of this feature we present the two dimensional projected density distribution of the halo and its surroundings in Fig. \ref{Fig:g15784density}. The density distribution appears smooth and nearly spherical within 0.6$R_{vir}$ (the inner dashed circle), while there is a distinct asymmetry near the virial radius (the solid circle). Beyond 1.5$R_{vir}$ (the outer dashed circle), the shape of the dark matter distribution follows the large-scale structure. Thus, the periphery of the dark matter halo is a non-spherical region at the interface of the spheroidal inner regions and the filament. The bulk angular flows of the gas are very much reduced inside the virial radius compared to the outside. It is noteworthy that there are strong angular flows exterior to the virial radius but weaker radial flows \citep{Prada2006} inside R$_{vir}$. 

The high density feature close to the virial radius of  g15784, which produces the sphericity dip, is at the interface of the two counter-rotating flows. Where they meet, the dark matter particles build up, creating the non-spherical and prolate feature. Inside the virial radius, rotation is positive and uniform. 

\begin{figure*}
\centering
     \begin{tabular}{cc}
     \includegraphics[scale=.35, trim={0.cm 1.5cm 1.4cm 3cm},clip]{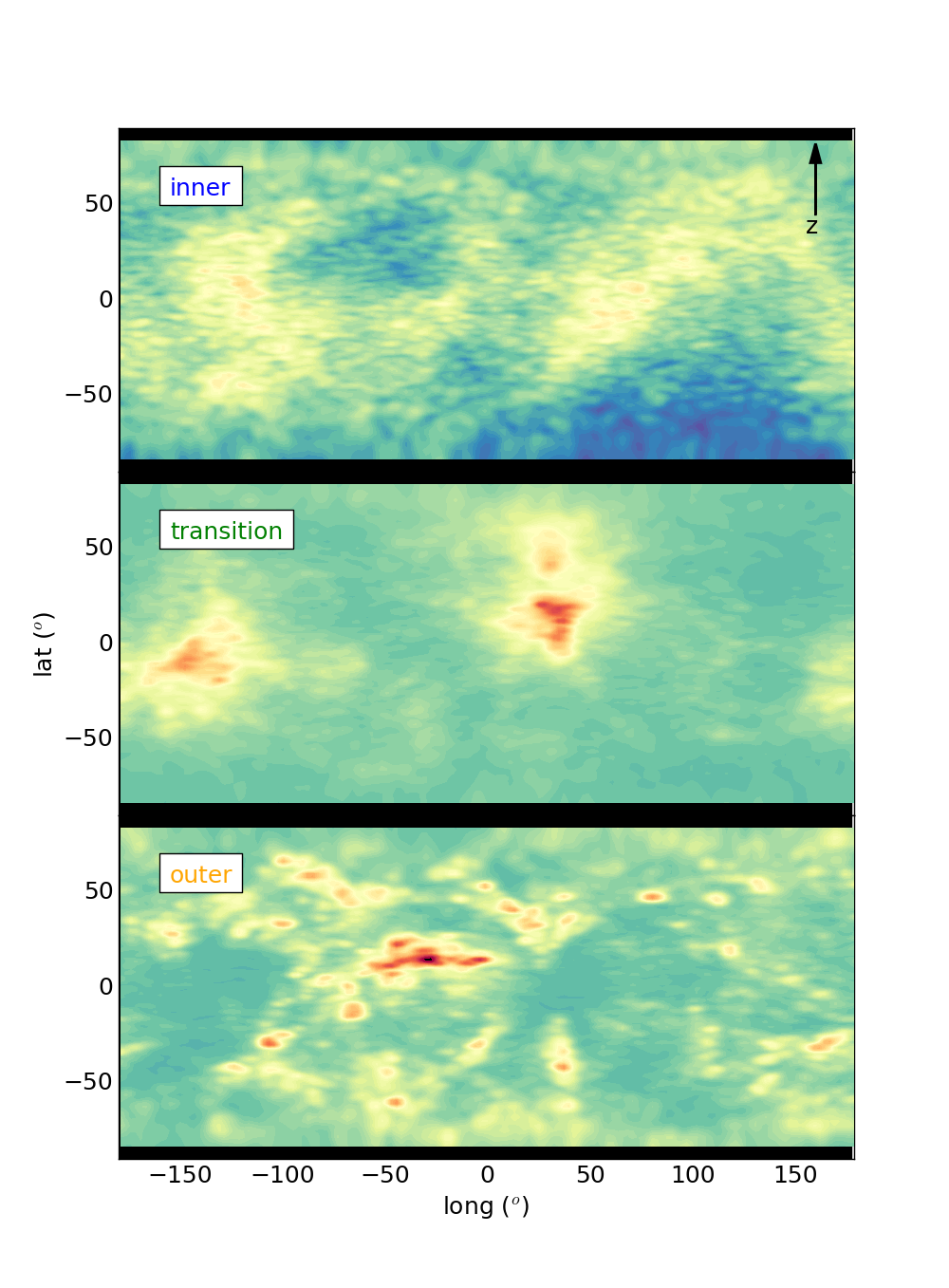} &
     \includegraphics[scale=.35, trim={0.cm 1.5cm 1.4cm 3cm},clip]{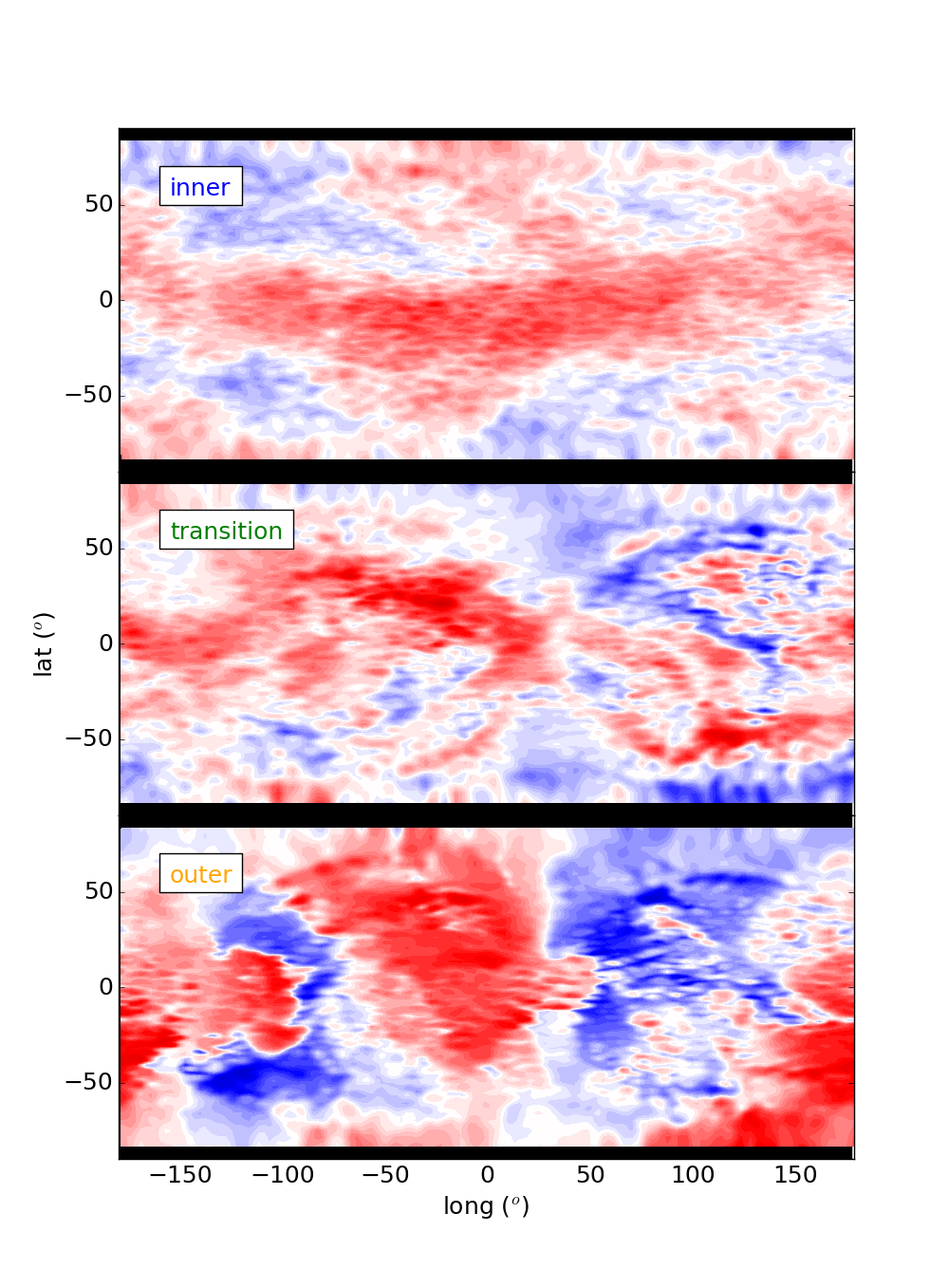} \\     

     \includegraphics[scale=.35, trim={0.cm 5.cm 1.4cm 5cm},clip]{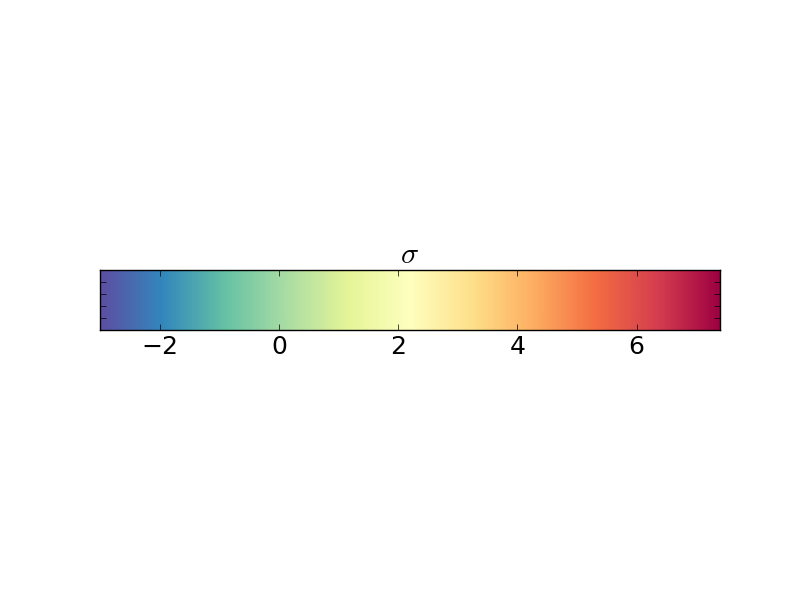} &
     \includegraphics[scale=.35, trim={0.cm 5.cm 1.4cm 5cm},clip]{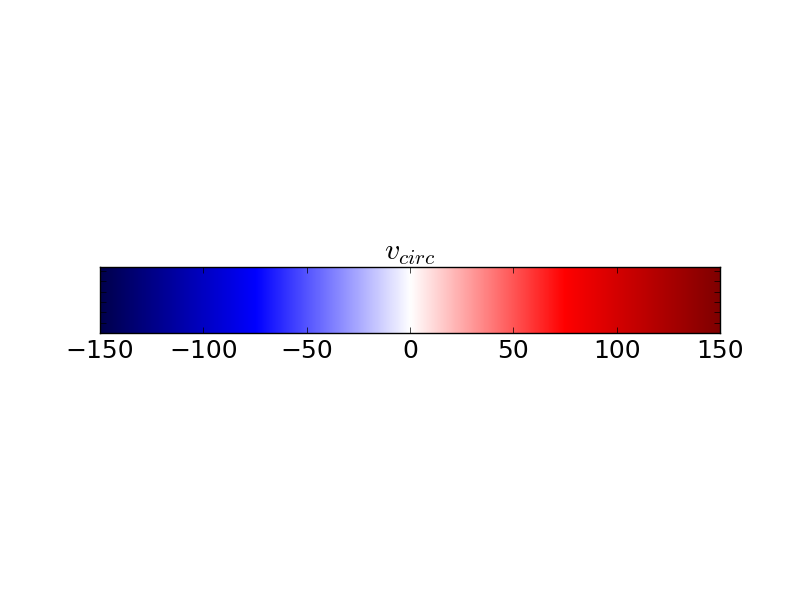} \\     
     \includegraphics[scale=.4, trim={0.cm 0.cm 0.0cm 0cm},clip]{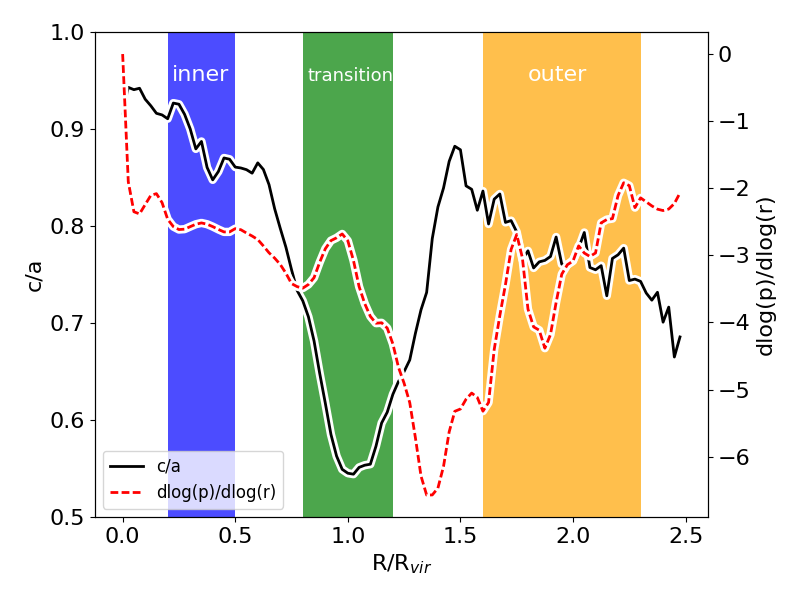} \\
     
     % trim={<left> <lower> <right> <upper>}               
\end{tabular}
\caption{The distribution of dark matter density (left) and circular velocity (right) in and around the dark matter halo of g15784. The top three panels in each column show the line of sight density distribution of dark matter density and circular velocity binned by longitude and latitude in different spherical shells, and then smoothed using a Gaussian kernel. The data are oriented so that the angular momentum of the galactic disc is aligned with the z-axis, and the panels are arranged in order of increasing radius. The upper three panels on the left are coloured in units of standard deviations ($\sigma$) from the mean of that panel, and in the right panel the colours indicate v$_{virc}$. The lower left panel shows the width of the spherical shells, and the z=0 sphericity and density gradient profiles for this halo. This panel shows the boundaries of radial regions, and the corresponding sections of the sphericity and density gradient profiles, shown in the upper panels. }
\label{Fig:mugsradialshells}
\end{figure*}

Fig.~\ref{Fig:mugsradialshells} shows mock `sky maps' of the dark matter distribution within spherical shells located inside, at, and beyond the virial radius. Beyond 1.6 $R_{vir}$, the density field is dominated by filamentary structures spread over the entire surface of the shell. Similarly, inside the virial radius, the density distribution is uniform, and relaxed. At the virial radius, the many filamentary structures visible at large radius have reduced to two: the regions of high density also visible in Fig. \ref{Fig:g15784density}. This implies that the region characterised by the sphericity dip and recovery is an often prolate transition region between the large-scale structure and the relaxed halo. Similarly, Fig. \ref{Fig:mugsradialshells}, right column, shows the sky map of the circular velocity in the x,y plane. Inside the dip feature there is differential rotation parallel to the gas disc, during the dip the velocity distribution is very irregular, while outside the dip a pronounced quadrupole can be seen caused by material flowing in from the filament. In the centre panel the quadrupole is clearly beginning to form and is fully developed by the 'outer shell'. The bottom panel of Fig. \ref{Fig:mugsradialshells} shows the sphericity and density gradient profiles of the halo,  and  illustrates the radial ranges used to generate the maps in the upper panels. This object does not show the two caustic feature of the average low $\Gamma$ halo, and the postion of splashback radius (1.4 R$_{vir}$) is slightly inside the average. However, some scatter is expected. We expect the two profiles to be linked, because the dip in sphericity may correspond to the second-to-last shell of the phase space distribution of the halo. Thus, the splashback and virial radii, marked by the sphericity dip, may be linked.

The importance of this feature is that it persists for a long time (8 Gyr), grows with the virial radius and is clearly evident in the density distribution. From this, we can conclude that {\it in the case of g15784} the virial radius is a clear interface between the dark matter halo and its environment, and its growth at late times is not merely pseudo-evolution due to the definition of the halo overdensity.

\subsubsection{Analysis of the full MUGS sample}

\begin{figure*}
\centering
\begin{tabular}{lll}
\multicolumn{3}{l}{\includegraphics[scale=.5, trim={0.cm 0 0cm 0cm},clip]{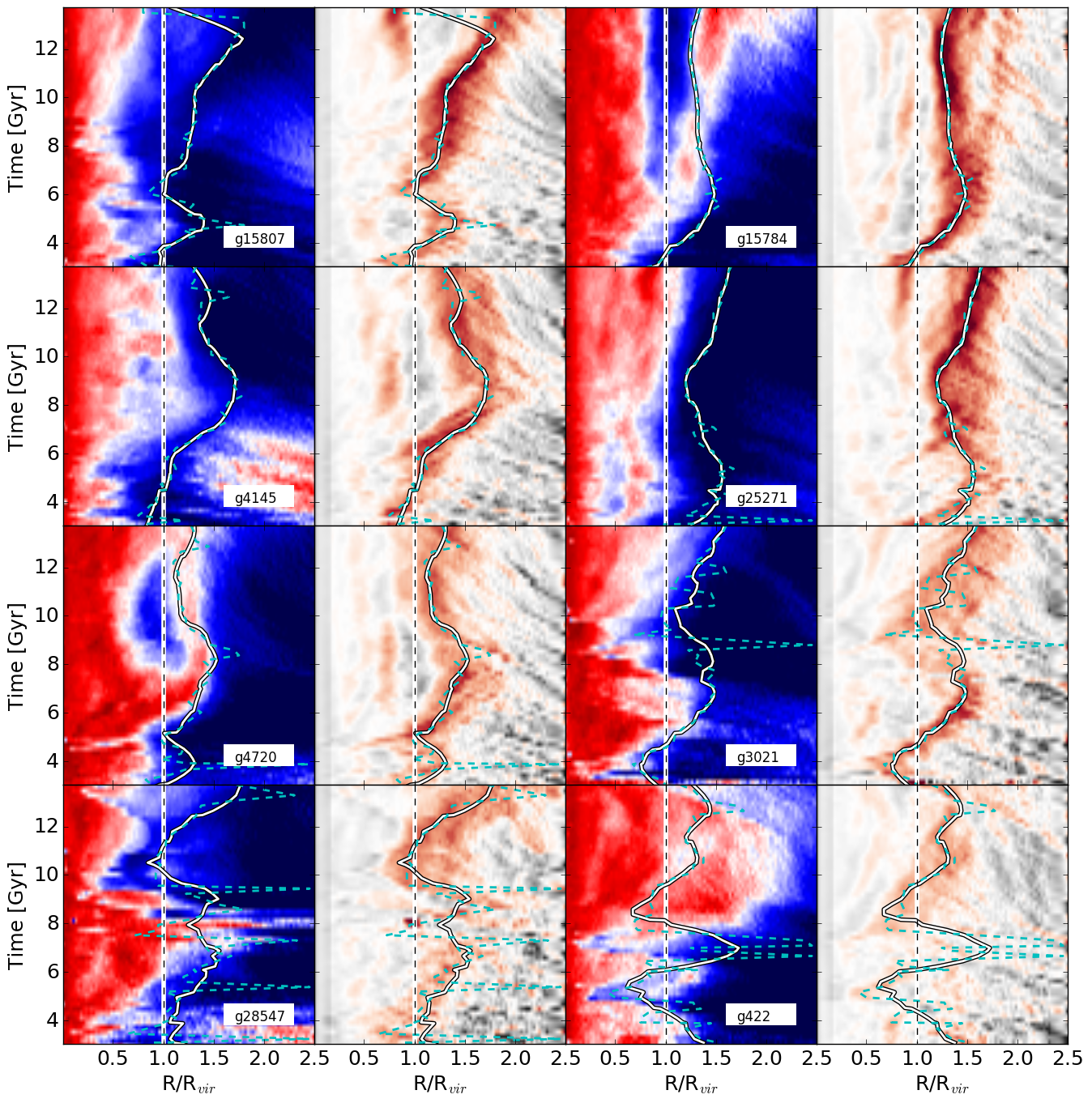}} \\
\includegraphics[scale=.3, trim={0.cm 2cm 0cm 5cm},clip]{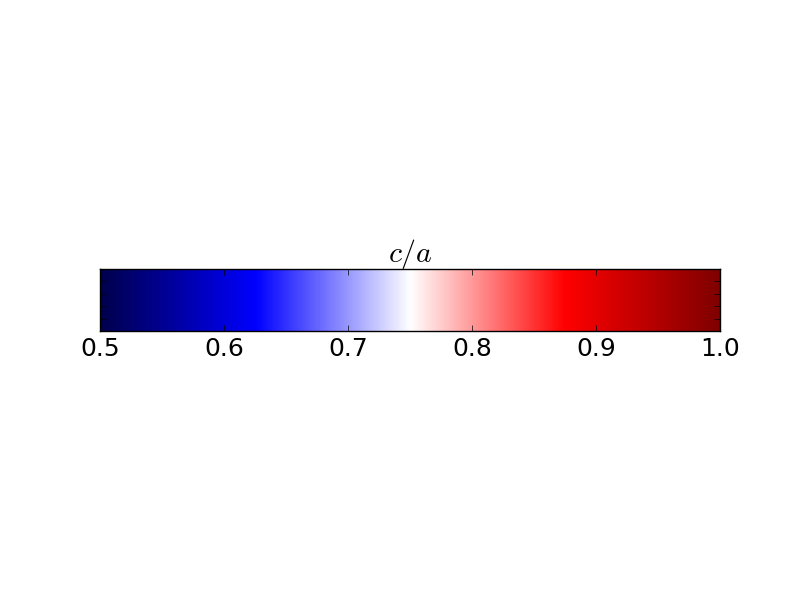} & & 
\includegraphics[scale=.3, trim={0.cm 2cm 0cm 5cm},clip]{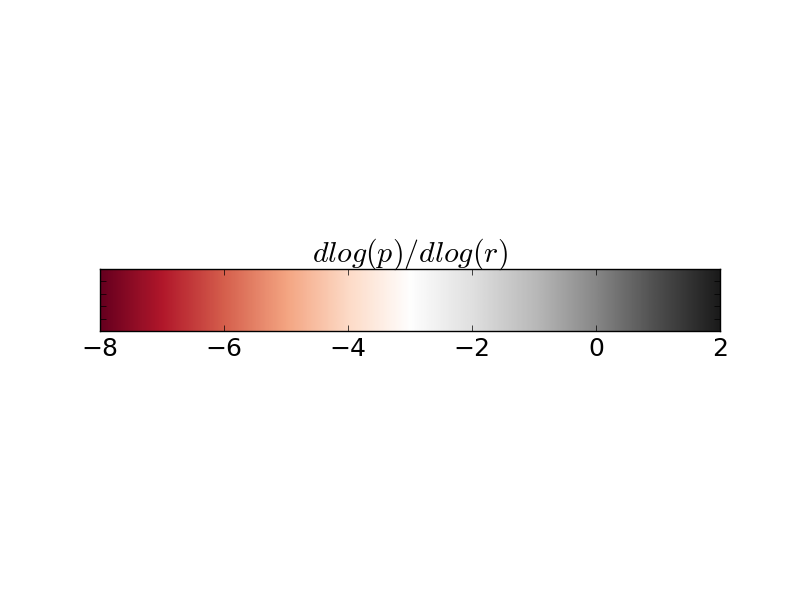}\\
\end{tabular}
\caption{Sphericity evolution for the MUGS galaxy sample (columns 1 and 3) for the 8 most massive halos in MUGS in order of virial mass. The galaxy label is given in each panel. The dashed line shows the virial radius and the white line is the smoothed splashback radius (unsmoothed is the dashed cyan line). The splashback radius is defined as the minimum of the density gradient in each time step. The density gradient map for the corresponding halo is shown in columns 2 and 4. The vertical lines at small radii  in columns 2 and 4 are an artifact of the smoothing method. There is a close relationship between the splashback region and the drop in sphericity where the halo starts to blend into the large-scale structure. }
\label{Fig:sphfeaturemugs}
\end{figure*}

\begin{figure*}
\centering
\begin{tabular}{lll}
\multicolumn{3}{l}{\includegraphics[scale=.5, trim={0.cm 0 0cm 0cm},clip]{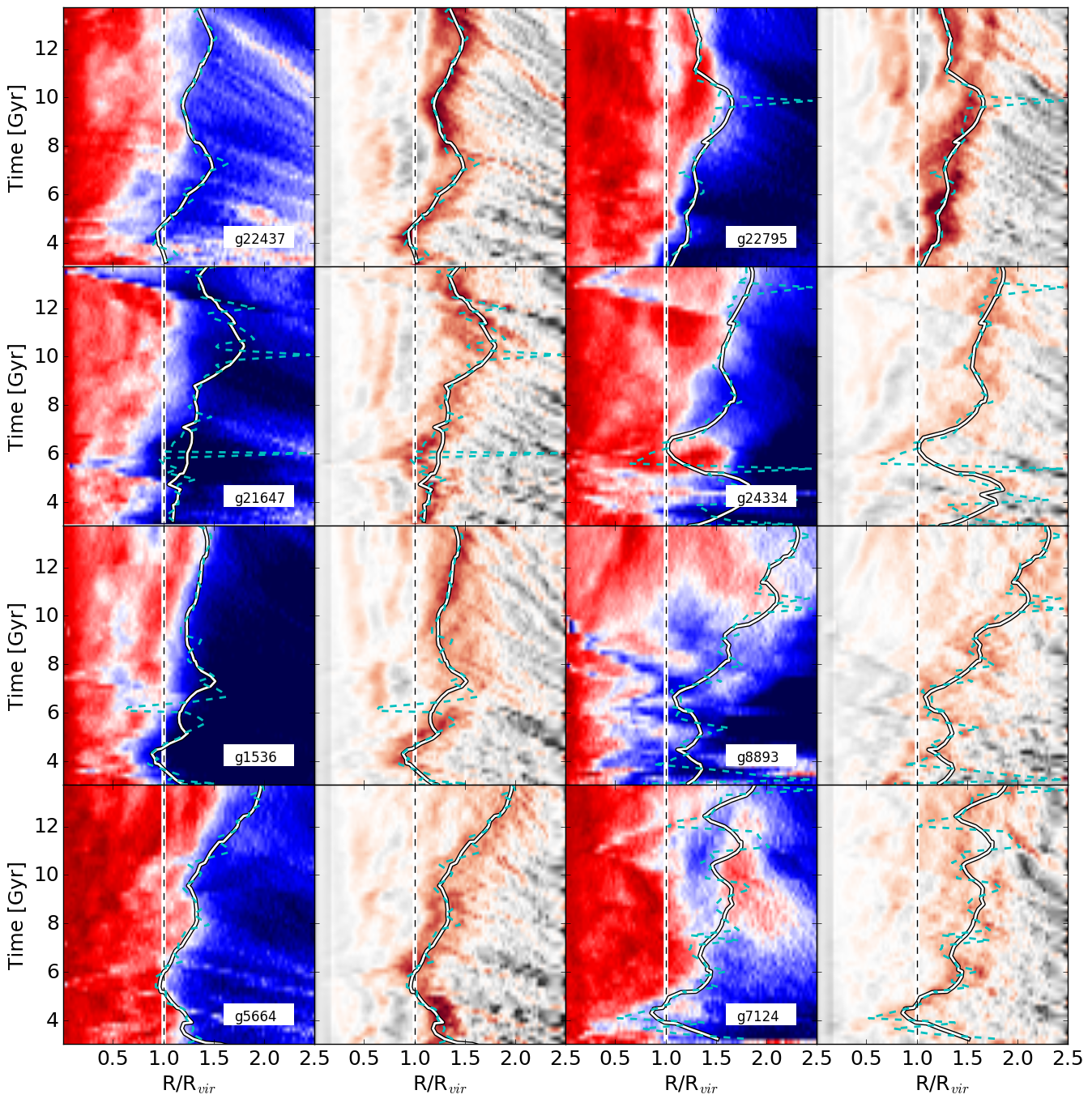}} \\
\includegraphics[scale=.3, trim={0.cm 2cm 0cm 5cm},clip]{compare_sph_with_splash_withsbr_cbar_csph_2.png} & & 
\includegraphics[scale=.3, trim={0.cm 2cm 0cm 5cm},clip]{compare_sph_with_splash_withsbr_cbar_cgrad_2.png}\\
\end{tabular}
\caption{Same as Fig. \ref{Fig:sphfeaturemugs} except for the 8 least massive halos in MUGS in order of virial mass.}
\label{Fig:sphfeaturemugs2}
\end{figure*}

It is difficult to generalize using a single galaxy. The principle advantage of the MUGS sample is that it contains a number of galaxies with a range of Hubble types \citep{Nickerson2013, Stinson2010} despite the limited mass range. Therefore, we can use the full sample of 16 galaxies to determine the general prevalence and origin of the shape feature at (or close to) the virial radius. We can, however, see hints of the general trends for larger ensembles.

 We have called the  valley in the density gradient the `splashback region', the minimum value of which is the splashback radius. This is shown as a coherent red region in Figs. \ref{Fig:sphfeaturemugs} and \ref{Fig:sphfeaturemugs2}. The splashback region is a striking, persistent feature and evolves over time. The minimum value of the density gradient is prone to significant scatter from snapshot to snapshot, as can be seen in the dotted line for halos g3021 and g28547, for example. In g3021 the minimum shifts from 0.5R$_{vir}$ to 2.5R$_{vir}$ in a few snapshots. This is due to noise or transient features. However, the splashback region itself is smoothly evolving, even if the minimum can spuriously change radius significantly (g15807 at z=0).  If plotted over time the splashback region is often much better defined than in a single snapshot. Some halos have stronger and deeper splashback regions than others, however. Although the sphericity dip is associated with the virial and splashback radii it is not `at' a particular radius. The processes which produce the dip are interpreted as a transition region between the environment and the halo proper.  When we attempt to identify the transition from the local environment to the halo proper, the splashback radius is found to be a consistently identifiable feature of a halo which evolves smoothly with time \citep[e.g.][]{Diemer2017a,Diemer2017b}. This is contrasts with the radius of the minimum of the density gradient profile, which suffers from a large scatter from snapshot-to-snapshot.
 
By plotting the evolution of the halo density profile with time it is easier to distinguish the region where the splashback is important. Many of the halos in MUGS show strong, well defined splashback regions which last for several Gyrs. Others, such as g24334, g8893 and g7124 are less well defined. The whole splashback region is obscured if substructure is included, and is very difficult to identify, particularly the outer edge. In order to identify the splashback region in individual halos it is vital to remove substructure, even in the case of g25271 for example, where the splashback region and splashback radius is well defined. 
 
 The splashback region in g15784 moves outwards with the increasing sphericity after the dip. This suggests a link between the processes which take place at the shape feature and the splashback and virial radii (after 8 Gyr). In this halo, matter close to the splashback radius is more spherical than the region just inside it.  This  suggests a link between the splashback region and the shape profile. The splashback region also appears to trace the outer edge of the most prolate region outside the virial radius. There appears to be a correspondence between the recovery and the splashback radius, but only at later times when the dip has become established. There is clearly, however, a non-trivial interaction between the halo shape evolution and the splashback region.
 
 In many cases there is a definite correspondence between the time evolution of the halo shape profile and the time evolution of the splashback region, suggesting that the splashback is an important driver of the halo shape. 

The presence of substructure in the host halo can change the shape of the halo locally. For example, the shape evolution of  g4720 is less clear when substructure is included but still clearly present (see Fig. \ref{Fig:examplewithsh} for a version of the 9th panel in Fig. \ref{Fig:sphfeaturemugs} which includes substructures). The shape feature is due to the entire local density field, which shows more variation when the high density (sub)halos are included. Substructure is transient, and does not persist at the same radius for long periods. Thus it is unlikely to be the cause of the dips where the dips are long lasting and visible, massive, substructures cannot be seen. The presence of substructures hides the shape of the global density field so we remove them. Strong, long term features seen in MUGS are robust to the presence of satellites.

\begin{figure*}
\centering
     \begin{tabular}{@{\hskip0pt}c@{\hskip0pt}c@{\hskip0pt}c@{\hskip0pt}c}
     \setlength{\tabcolsep}{0pt}
     g15784 & g1536 & g28547 & g15807 \\
     \includegraphics[scale=.3, trim={0.cm 1.cm 0.cm .0cm},clip]{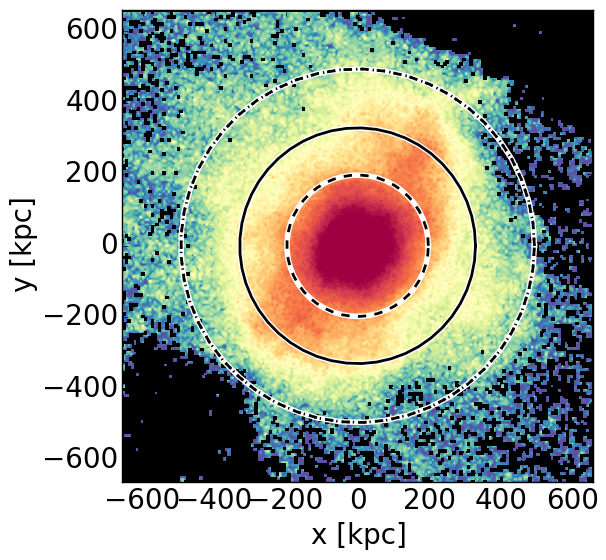} &
      \includegraphics[scale=.3, trim={1.cm 1cm 0.cm .0cm},clip]{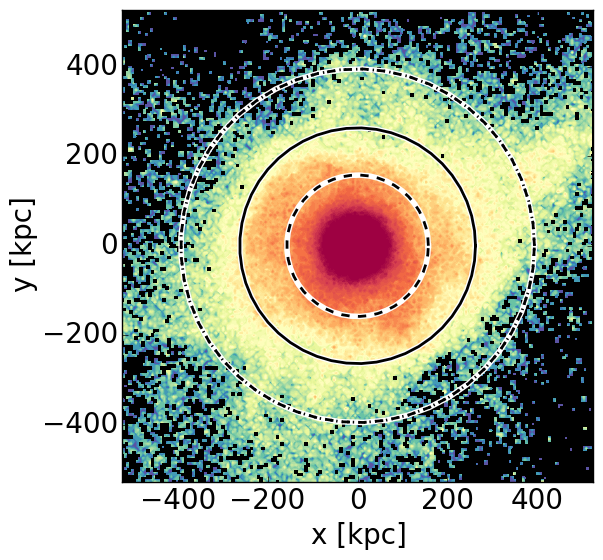} &
      \includegraphics[scale=.3, trim={1.cm 1cm 0.cm .0cm},clip]{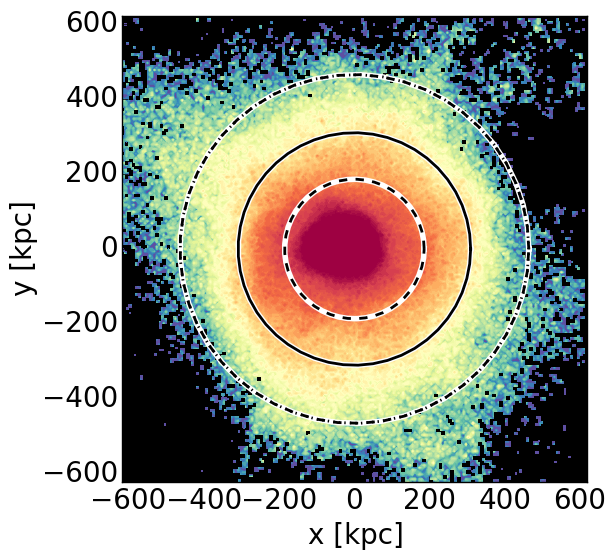} &
      \includegraphics[scale=.3, trim={1.cm 1cm 0.cm 0.0cm},clip]{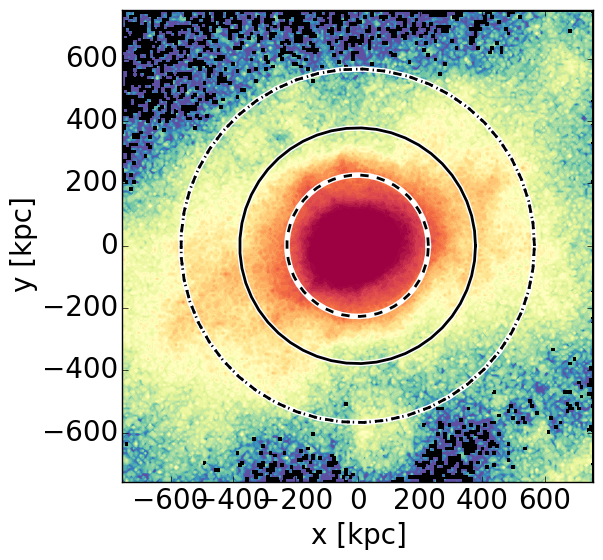} \\
      
     \includegraphics[scale=.3, trim={0.cm 1.cm 0.0cm .0cm},clip]{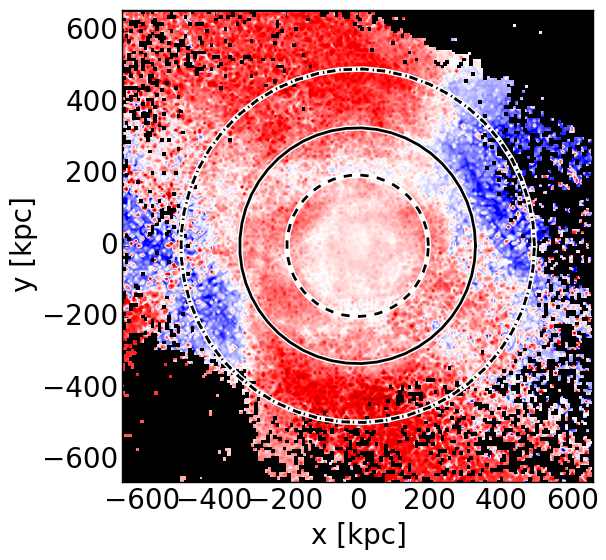} &
      \includegraphics[scale=.3, trim={1.cm 1cm 0.0cm .0cm},clip]{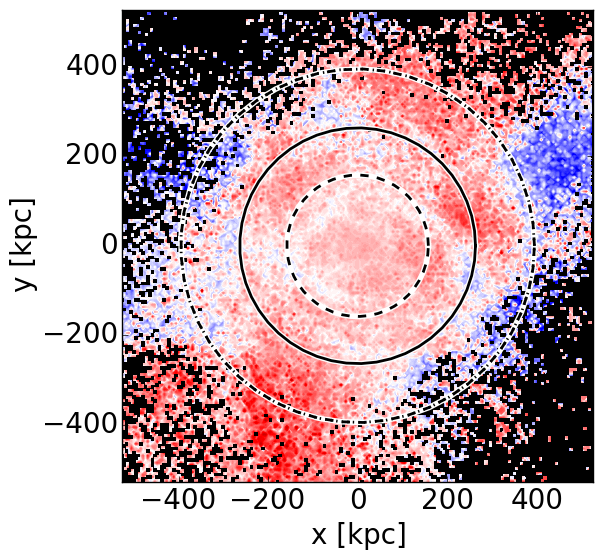} &
      \includegraphics[scale=.3, trim={1.cm 1cm 0.0cm .0cm},clip]{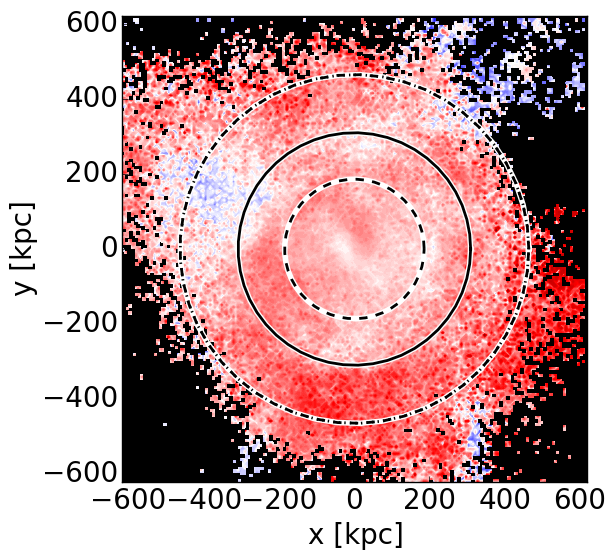} &
      \includegraphics[scale=.3, trim={1.cm 1cm 0.0cm 0.0cm},clip]{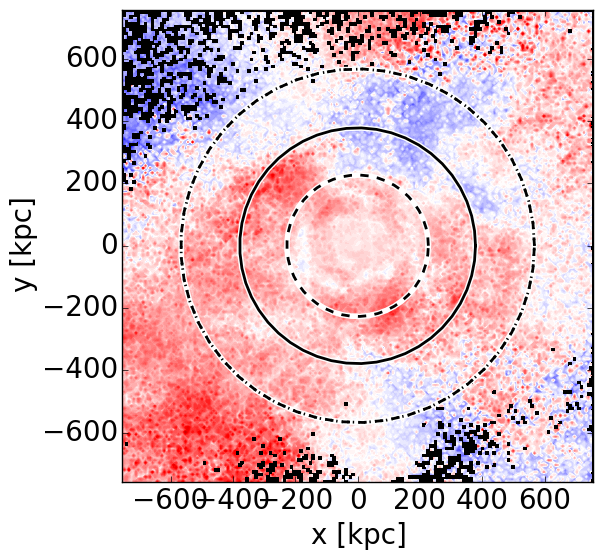} \\   
      
     \includegraphics[scale=.3, trim={0.cm 1.cm 0.0cm .0cm},clip]{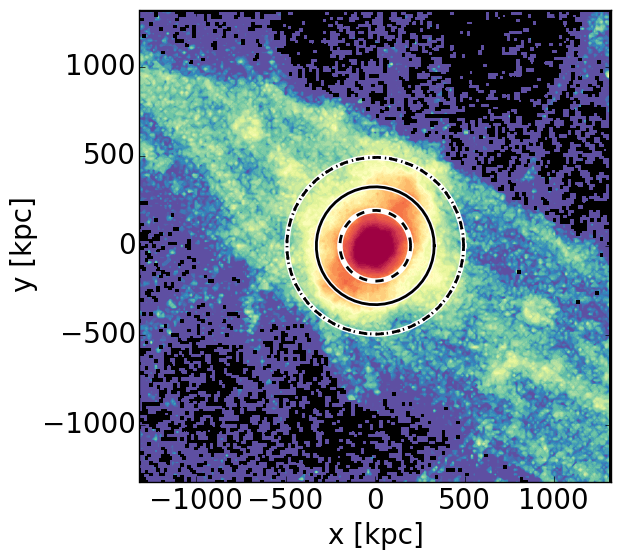} &
      \includegraphics[scale=.3, trim={1.cm 1cm 0.0cm .0cm},clip]{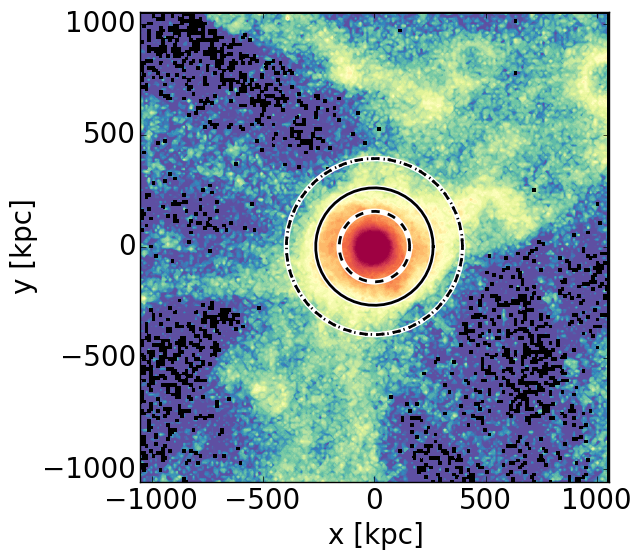} &
      \includegraphics[scale=.3, trim={1.cm 1cm 0.0cm .0cm},clip]{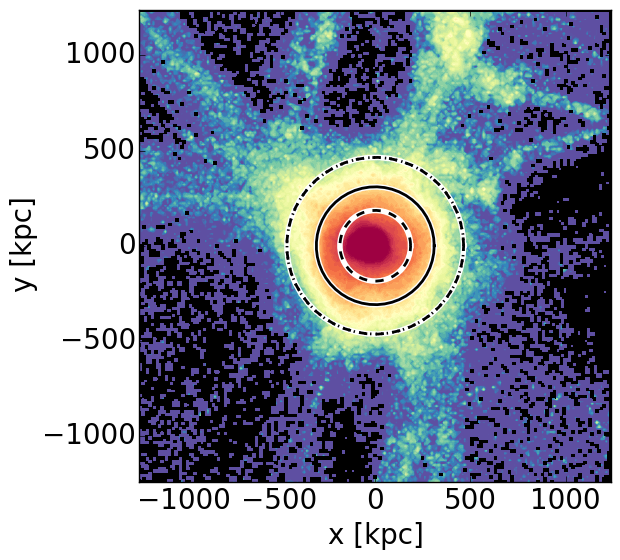} &
      \includegraphics[scale=.3, trim={1.cm 1cm 0.0cm 0.0cm},clip]{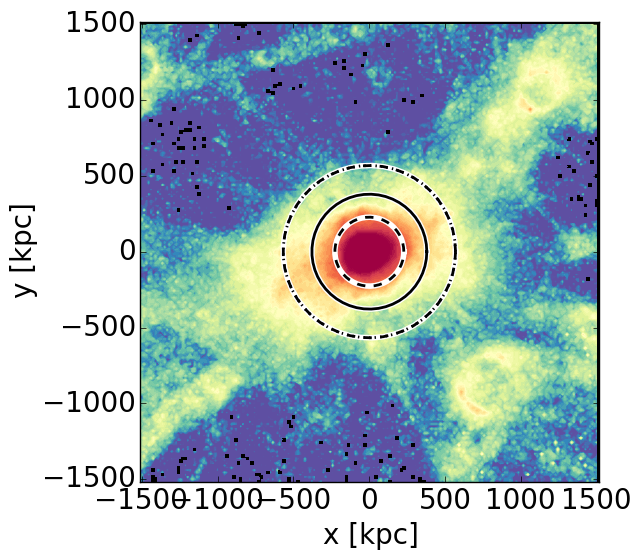} \\
      
     \includegraphics[scale=.3, trim={0cm 0.cm 0cm .0cm},clip]{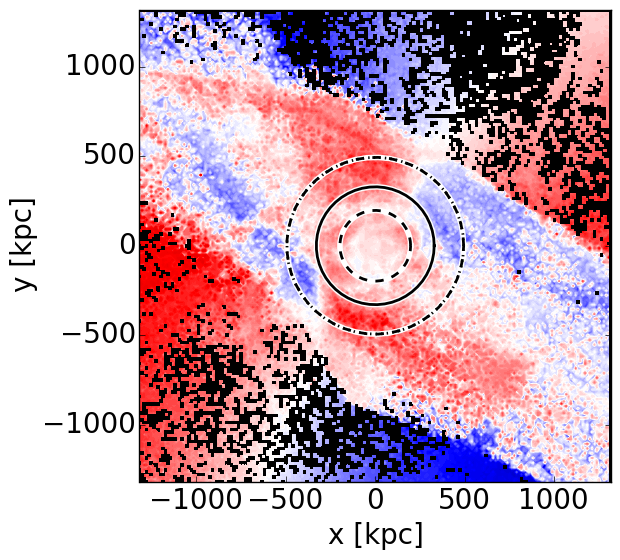} &
      \includegraphics[scale=.3, trim={1cm 0cm 0cm .0cm},clip]{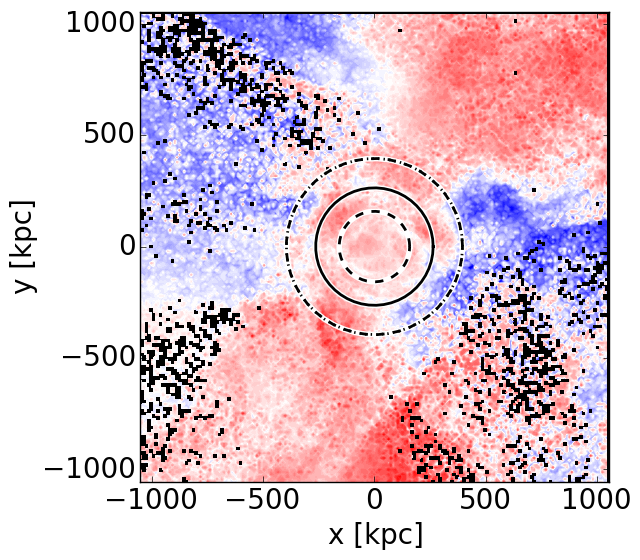} &
      \includegraphics[scale=.3, trim={1cm 0cm 0cm .0cm},clip]{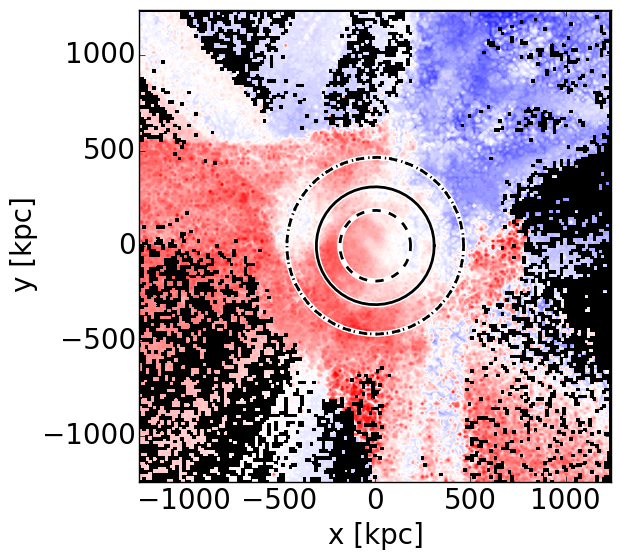} &
      \includegraphics[scale=.3, trim={1cm 0cm 0cm 0.0cm},clip]{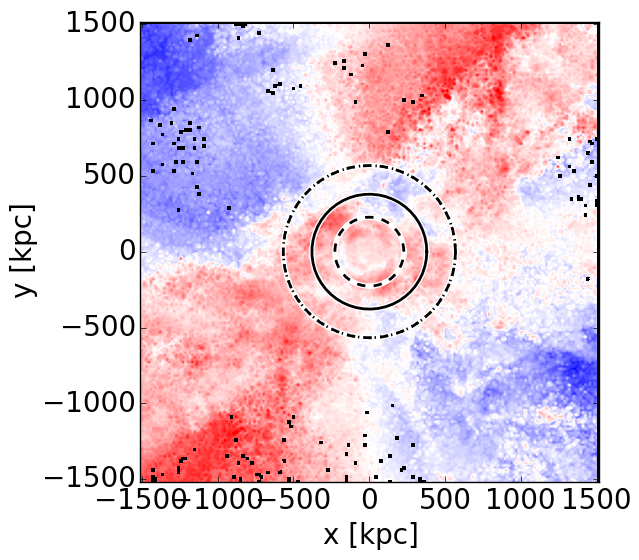} \\            
      
     % trim={<left> <lower> <right> <upper>}               
\end{tabular}
\caption{Top row: The distribution of dark matter in and around the dark matter haloes of g15784, g1536, g28547 and g15807 (left to right) within a cube of side length $\pm$2$R_{vir}$. Second row: The angular velocity of dark matter of these same four galaxies. Third row: The distribution of dark matter in and around the dark matter haloes of the four galaxies in a cube of side length $\pm$4$R_{vir}$. Second row: The angular velocity of dark matter of these same four galaxies. The dashed, solid and dot-dashed lines show 0.6, 1 and 1.5 times R$_{vir}$. Halos g15784 and g1536 show strong dips, g28547 shows no dip and g15807 shows a plateau at the virial radius. Where the dips are strong coherent infall can be seen, while where the dip is not apparent infall is more spherical.}
\label{Fig:mugsdistributions}
\end{figure*}

To investigate whether this correspondence applies to other halos, we explored the angular velocity and density maps of the subset of haloes which show a clear shape feature close to the virial radius (g15784, g4720, g22795, g1536  and g7124) and the maps of some halos which do not (see Fig.~\ref{Fig:mugsdistributions}). Where the dip is strong we see a quadrupole in the angular velocity distribution, where the dip is weak, or not present, this feature is less strong.  Within the MUGS haloes with the strong dip, the counter-rotating flows settles out quickly and by 0.6 R$_{vir}$ there is very little structure in angular velocity, and the shape profile is mostly spherical. This implies that there is a considerable subset of halos which experience angular restructuring close to the virial radius. 

In general there are three sections in a dark matter halo and its local environment, although these regions are given only as a rough guide due to the large halo-to-halo variations: 

\begin{enumerate}
\item $R<R_{vir}$: the dark matter distribution is relaxed, smooth and approximately spherical.
\item $R_{vir}<R<R_{splash}$: contains both inbound and outbound material and is clearly within the region of influence of the halo. However, the dark matter is clumpy and unrelaxed.
\item $R>R_{splash}$: Outside the splashback radius is the `large-scale structure' region which has never been affected by the halo.
\end{enumerate}

The processes that give rise to the shape distribution (and density gradient profile) in these regions are different, and so it is often the case that they show different shapes, the details of which depend on the geometry. For example, if four filaments feeding into the transition region are in a plane, the resulting density distribution will favour an oblate ellipsoid, while if those filaments are arranged in a  tetrahedron then a sphere will be preferred.

Figures \ref{Fig:sphfeaturemugs} and \ref{Fig:sphfeaturemugs2} demonstrate that the splashback radius can move inside the virial radius, particularly at early times. This is also expected when the rate of infall, $\Gamma$, is high \citep{Diemer2014}. In the Illustris sample the strongest dip in sphericity at high $\Gamma$, where the radius of redistribution is driven inside the virial radius e.g. Fig. \ref{Fig:Illustrisbinedsplash}. The over-correction outside the dip is in a similar location at 1.2 R$_{vir}$ at all $\Gamma$. In the case there R$_{splash}<$R$_{vir}$ the redistribution of material takes place inside the virial radius. This is logical, considering that at very high infall rates the halo is disturbed in the outer regions and not relaxed, while relaxation can still occur further in.

Figure \ref{Fig:mugsdistributions} hints at the importance of the geometry of the system. The local environment of the halo appears to have an influence on the velocity field. The lower two rows of the figure suggest that a strong and {\it coherent} filament, as in g15784, produces a strong quadrupole, while a weaker, multi-directional inflow, as in g28547, produces no shape feature.  However, even in this case, inspection of the density distribution suggests that the virial radius roughly marks the transition between the smooth inner region and the filamentary outer region.

If $R_{vir}$ and $R_{splash}$ are close to each other then the second region can be fairly narrow, and so there is a narrow region close to $R_{vir}$ that has a very distinct shape. This manifests most dramatically in cases like g15784 and the most massive Illustris halos, shown in Fig. \ref{Fig:Averages}.

\section{Discussion \& Conclusion}\label{discuss}
We identify a strong correlation in the shape of the density distribution around dark matter halos which lies roughly between the splashback and the virial radii. This can be seen in the `average' shape profile for massive halos in the Illustris simulation, and in individual cases, using the MUGS zoom simulations. In many cases (but not all) the splashback radius lies  along the region of the shape profile that sees a long term drop in  sphericity as the outer region of the halo mixes with the large-scale structure. There is a correspondence between the splashback region and a `recovery' of the sphericity profile to higher sphericity. This implies that there is important information encoded in the second order moments of the density field. 

We find a relation between the outer edge of the spherical region of the halo, and the `splashback region' close to the splashback radius  \citep[e.g.][etc.]{Diemer2014, More2015}. Sometimes this splashback region coincides with the recovery of the density field, but is often close to the interface between the halo and the non-spherical large-scale structure.

In the Illustris simulation there is evidence that the dip in the shape profile is most evident when the splashback radius is close to the virial radius, although the same thing is not clear in the MUGS simulation for individual galaxies due to strong object-to-object variation. This implies, contrary to work on the density profile of galaxies beyond the virial radius \citep{Diemand2007,Diemer2013,Cuesta2008}, that the virial radius has a physical meaning in terms of the properties of galaxies, {\it at least in a significant subset of dark matter halos.} The density gradient profile shows an inner dip close to the virial radius which corresponds to the inner caustic of the density field \citep{Adhikari2014}. This suggests that the shape feature is related to infalling material having had more orbits to virialize inside R$_{vir}$.  In spherical collapse model presented in \citet{Adhikari2014} the inner caustic corresponds to the second orbit of material. This is also true in MUGS where material at the inner density gradient dip is at its second orbit, where it is better mixed with the in-situ material.

We observe that the edge of the dark matter halo, defined by the virial/splashback radius, coincides with the dominance of the quadrupole in the circular velocity distribution, indicating that material flowing into the halo from the large-scale structure is rapidly mixed in the dark matter halo.  We can see that bulk flows do occur near the halo edge, but these are angular flows rather than radial. We see a net re-distribution of material at any given radius, but the radius of that material does not change. The matter is redistributed from the direction of the filament to almost  90$^o$ to it.   The splashback radius defines roughly where material has reached the apocentre of its first orbit, while, we conjecture, the virial radius defines a boundary where the angular momentum redistribution has time to act, with the dip coinciding with the apocentre of the second orbit\footnote{This could be followed up by tracing the orbits of particles at different radii.}. These effects influence the shape profile, seen in both individual galaxies and in aggregate. This can manifest as a dip in sphericity close to R$_{vir}$.

The region roughly between the splashback and virial radii marks a transition zone between the environment and the halo. Thus,  a potential explanation for our results is that inside the {\it virial radius} the halo is virialised and phase mixed, outside the {\it splashback radius} material is filamentary and exhibits the properties of the large scale structure. Inside the splashback radius, and before reaching the virial radius, material is not wholly filamentary but remains unrelaxed. The region between the virial and splashback radius sees mixing between newly inflowing material, but this material has not had time to fully mix with the halo. This interpretation is supported by Fig. \ref{Fig:Illustrisbinedsplash} and Fig. \ref{Fig:splashcomp} where the dip is strongest where the accretion rate is high. In order for there to be a strong interaction between virializing and newly infalling material the accretion rate must be high and structured, and this is clearly shown in panel (b) of the figure.

This has also been seen by \citet{Suto2016}, who noticed the correspondence between halo shape and radial velocity dispersion.  {\it \citet{Suto2016} found that the splashback radius matched with the start of a fall in radial velocity dispersion} in the outer regions of halos, which extends out into the large-scale environment. This suggests the splashback radius marks the point where the virialized material starts to affect the profile, and that at $r>R_{splash}$ the large scale structure is dominant. As the virialized material tends to be more spherical, and the large-scale structure is filimentary, this is consistent with our scenario. However, further work is required to test the validity of this interesting possibility, e.g. \citet{Diemer2017a}, \citet{Mansfield2016} and \citet{Diemer2017b}.   Future work should explore the orbits of the particles outside, and inside  the transition region, and explore how their properties evolve with radius and time as they relax. 

In our interpretation, both radii are important because together they demarcate the inner and outer edges of the transition region, with some significant blurring on each side. In between, the density field is dominated by two clumps, due to the mixing of material infalling from the bulk flows in the large-scale structure. The evidence of the bulk flows in Fig. \ref{Fig:mugsradialshells}, right column, is less clear in the interface region, because some mixing has occurred. The material has not wholly relaxed (as at r$<$R$_{vir}$). Clearly, however, there is considerable halo-to-halo variation and evolution. This behaviour appears to hold true only in general, however, and individual halos can vary considerably. Rapid infall and interactions can disturb the outer layers of a halo, meaning that at larger radii the halo is not relaxed and not virialized. It could then be argued that the virialized region of the halo does not extend fully to the virial radius expected from the cosmology, but that the effective radius of virialization is smaller than this. This would make the inner edge of the transition region variable, because the inner edge is the radius at which material has effectively relaxed.

In Fig. \ref{Fig:Averages}, the sphericity dip corresponds with the density gradient dip produced by the inner caustic, and this inner caustic is produced by the second apocentre of infalling material \citep[e.g.][]{Adhikari2014}. Inside the inner caustic, material is well mixed and relaxed, outside the inner caustic material is in the process of splashback.  Thus, the sphericity dip marks the intermediate state between fully virialised material and material that maintains the coherent phase space distribution it had at infall. During its second orbit material is beginning to phase mix, but has not yet fully done so. That this radius should be in the vicinity of the virial radius is unsurprising, because in the spherical collapse model the virial radius marks the boundary at which material has the chance to relax.

 At this point we wish to speculate about the physical origin of the shape feature. Material flows into the halo along a preferential direction, and so has a non-circular orbit. The asphericity at the splashback radius is not apparent from our methods because there is an inherent asphericity in the background density field. At the inner caustic, material is in the process of relaxing, but has not fully phase mixed. There is a build up of material  at the second caustic because material is still on elliptical orbits and this produces the sphericity dip. The exact position, and shape, of the dip depends on the properties of the infall, beyond simply the infall rate, $\Gamma$, because there is a directional component. 

In g15784, for example, the two caustics are not apparent, but the dip is strong. This suggests that although the first and second apocentres cannot be distinguished, a strong ellipsoidal feature is produced by material falling in from the narrow and coherent filament in which this halo is embedded. This needs to be examined further by tracing the orbit of material through time  over multiple orbits \citep[i.e.][]{Diemer2017a}. 
 
The existence of a smoothly evolving (and in many cases strong) splashback region in dark matter halos suggests that even if the identification of the splashback radius in individual halos using the density gradient approach \citep[e.g.][]{More2015, Diemer2014} is difficult, the splashback region is often very well established for most of cosmic time.  The formal definition of the splashback radius as the minimum of the density gradient profile exhibits a strong degree of scatter from snapshot to snapshot (the dashed cyan line in Fig. \ref{Fig:sphfeaturemugs} and \ref{Fig:sphfeaturemugs2}). This is because spurious dips in the density gradient profile can exist for a brief interval, and then fade by the next snapshot. The splashback region is, in many cases (but not all) robust and long lived. Using individual simulation snapshots these spurious features are not trivial to identify, but the robust splashback region is evident using time evolution information.

The radius at which most of the material is virialised is not  the formal virial radius. This is expected from the non-spherical form of the overdensity contours produced by halo finders such as SUBFIND, and the non-spherical form of the splashback region \citep{Mansfield2016}.

When we discuss the importance of the virial radius as the inner edge of the transition region we are only being approximate. The actual transition region can occupy radii smaller than this depending on properties of infalling material. It it perhaps more useful to discuss the radius at which the majority of material is virialized (the effective virial radius). At the inner caustic this process is beginning and so inside the sphericity dip the halo is virialised. The dip demarcates the transition region and the radius of the second orbit where this transition is taking place.  

In summary, we suggest that because material at the inner caustic/second orbit retains some of the orbital information it had at infall there is a  tendency for material at this radius to have a non-spherical shape. This material is on elliptical orbits and is the direct cause of the dip. Inside this, material has had time to virialize, and is on more spherical orbits. The higher the infall rate and more elliptical is distribution, the deeper the dip. At this radius, the elliptical material is emphasised over the background of already virialized material. At higher $\Gamma$ the non-virialized material penetrates deeper into the halo and the outer regions of the halo are less spherical (e.g. Fig. \ref{Fig:Illustrisbinedsplash}). 

\section*{acknowledgements}
ONS and JB acknowledge support for program HST-AR-12837 was provided by NASA through a grant from the Space Telescope Science Institute, which is operated by the Association of Universities for Research in Astronomy, Inc., under NASA contract NAS 5-26555. ONS thanks the Korea Institute for Advanced Study (KIAS) Center for Advanced Computation Linux Cluster System for providing computing resources for this work. This work has made use of the Shared Hierarchical Academic Research Computing Network (SHARCNET) Dedicated Resource Project: ``MUGS: The McMaster Unbiased Galaxy Simulations Project'' (DR316, DR401 and DR437). AK is supported by the {\it Ministerio de Econom\'ia y Competitividad} and the {\it Fondo Europeo de Desarrollo Regional} (MINECO/FEDER, UE) in Spain through grant AYA2015-63810-P as well as the Consolider-Ingenio 2010 Programme of the {\it Spanish Ministerio de Ciencia e Innovaci\'on} (MICINN) under grant MultiDark CSD2009-00064. He also acknowledges support from the {\it Australian Research Council} (ARC) grant DP140100198. He further thanks Slowdive for souvlaki. This research made use of pynbody \citep{Pontzen2013}, the following python libraries: scipy \citep{Jones2001}, IPython \citep{Perez2007}, matplotlib \citep{Hunter2007} and numpy \citep{Walt2011}, as well as the tools made available by the Illustris collaboration for reading their data. We thank the referee for their comments, which have greatly improved the quality of the paper.

\bibliographystyle{mn2e}
\bibliography{biblio2}

\appendix
\section{Splashback radius}
\label{Append_slp}

\begin{figure}
\centering
     \begin{tabular}{c}

     \includegraphics[scale=.4, trim={0.cm 0.cm 1.7cm .0cm},clip]{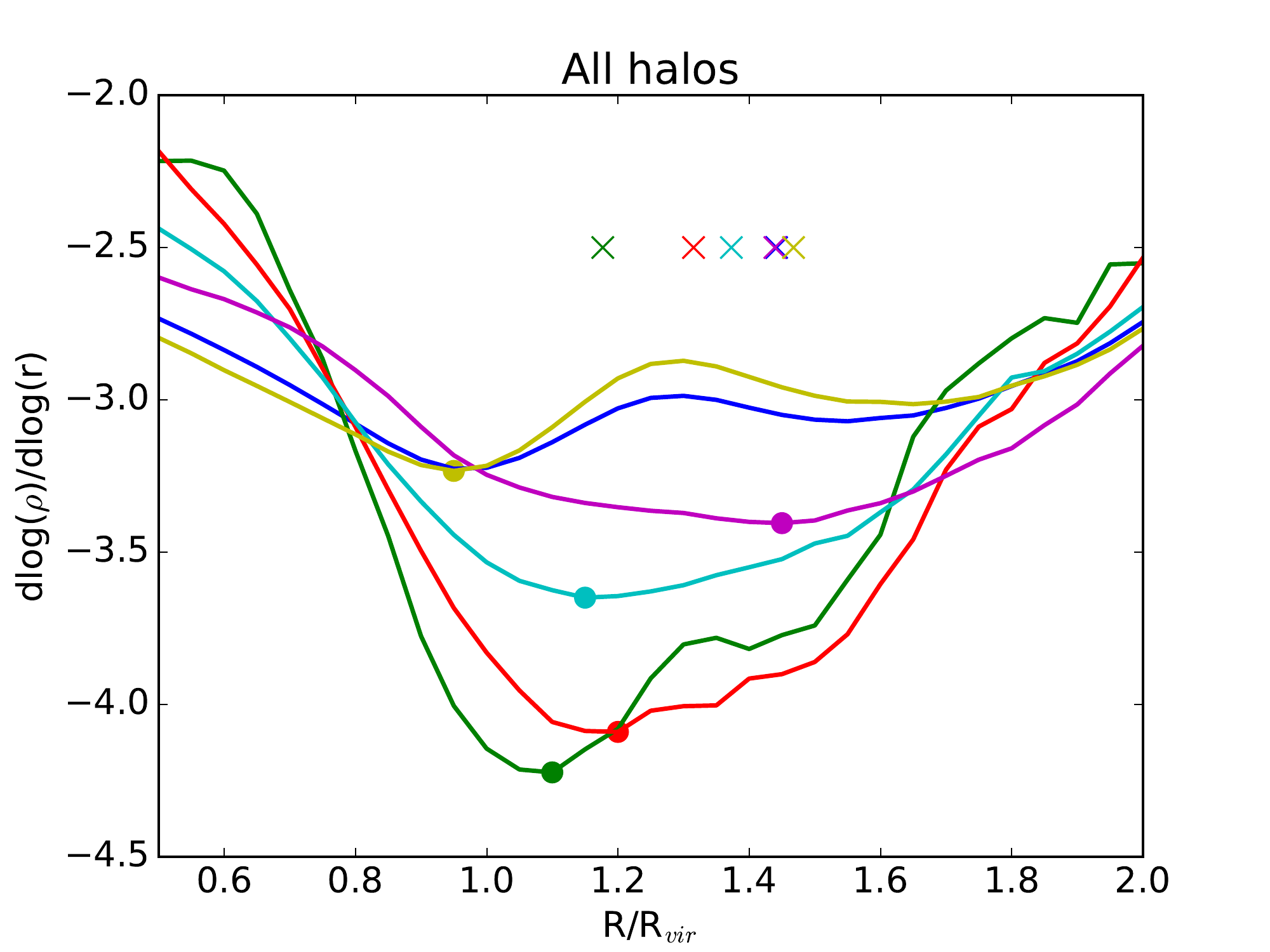} \\
     \includegraphics[scale=.4, trim={0.cm 0.cm 1.7cm .0cm},clip]{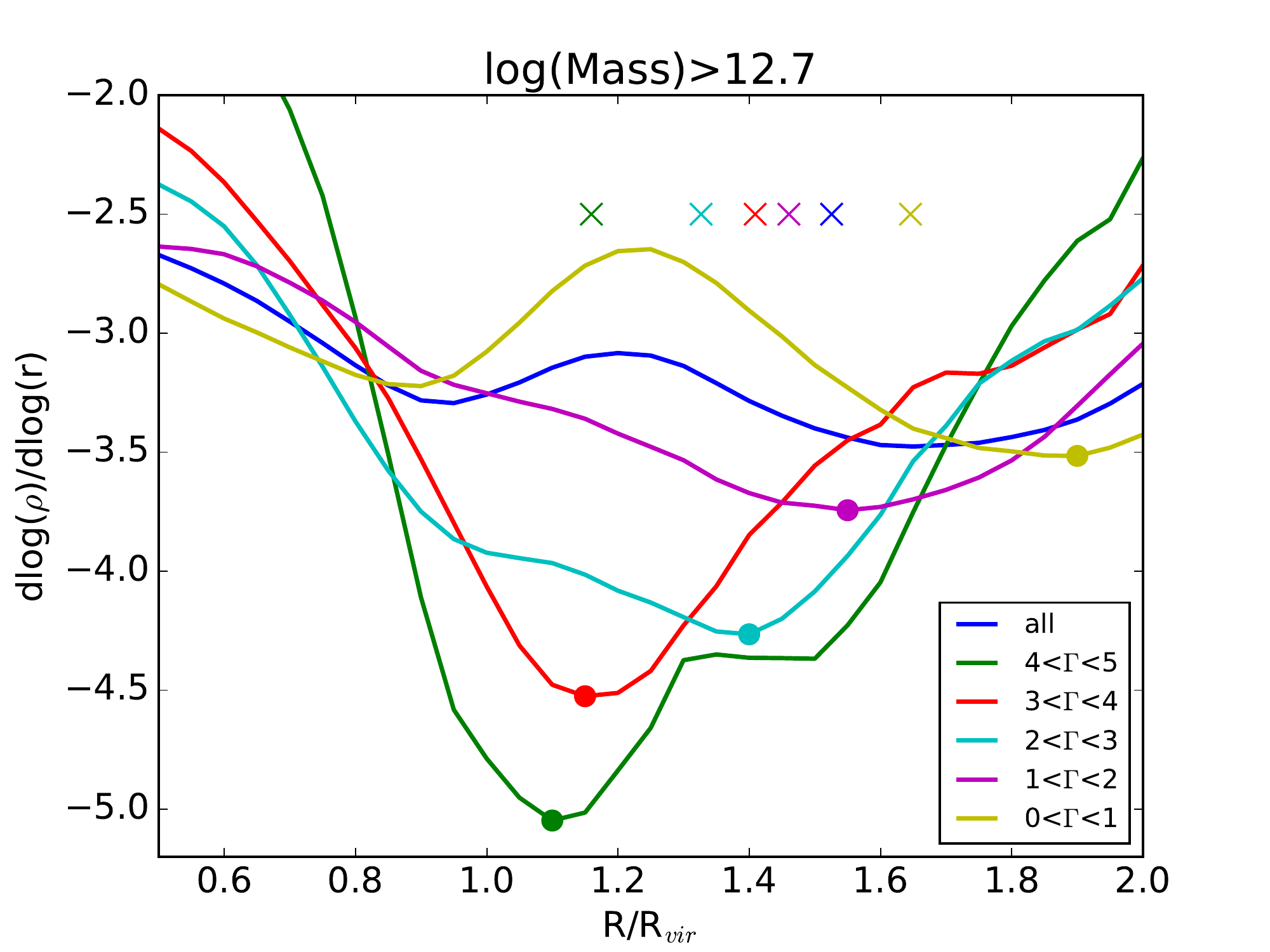} \\               
     \includegraphics[scale=.4, trim={0.cm 0.cm 1.7cm .0cm},clip]{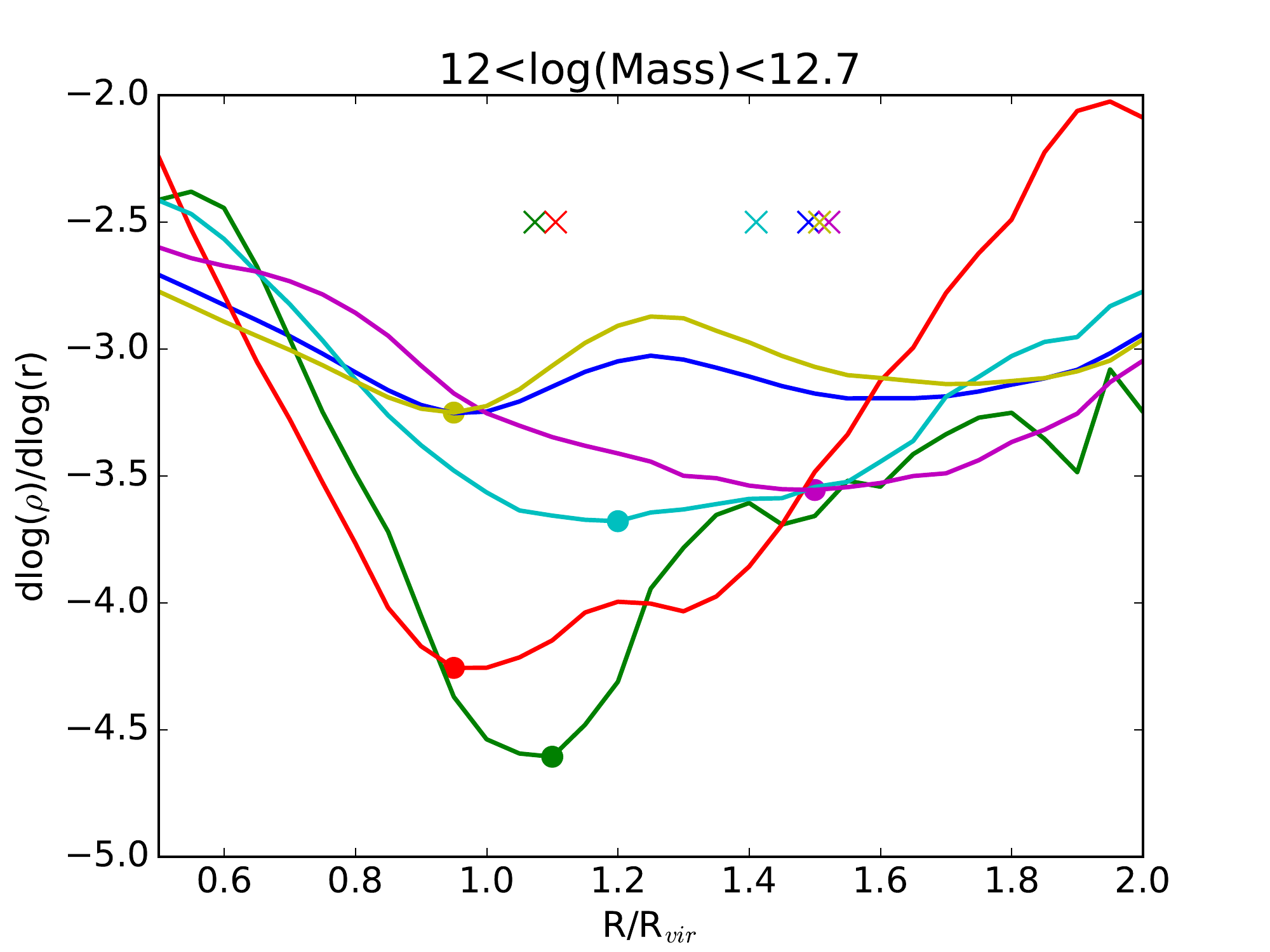} \\ 
               
     % trim={<left> <lower> <right> <upper>}               
\end{tabular}
\caption{The mean density gradient profiles for the Illustris-1-Dark sample. The top panel shows the mean profile of all halos, and the middle and bottom panels show the profiles for different mass bins. Each panel shows the profiles decomposed by infall rate as parameterised by $\Gamma$. The points indicate the recovered splashback radius. The crosses indicate the mean of the splashback radius calculated for individual halos. }
\label{Fig:splashtest}
\end{figure}

Our approach to identifying the splashback radii of halos differs from \citet{Diemer2014} and \citet{More2015} in that we find the averages of the density profiles of individual galaxies, excluding subhalos. A further difference is that our halo sample includes only  the dark matter particles making up host halos \citep[where we exclude subhalos, c.f.][]{Mansfield2016}. We also present mean profiles, rather than median profiles.

The two caustics shown in Fig. \ref{Fig:Illustrisbinedsplash} are  mainly due to the contribution from the halo sample with the lowest $\Gamma$. This is also seen in Figure 7 of \citet{Diemer2014} and Fig. 5,7 and 9 of \citet{Adhikari2014}, although the effect on the overall profile is not as strong. This is presumably due to the differences in the source halo sample, or the method used identify the average (i.e. the mean rather than the median) density gradient. The two caustic feature is strongest in high mass halos with low $\Gamma$, as can be seen in the middle panel of Fig.\ref{Fig:splashtest}. The smoothly increasing splashback radius with $\Gamma$ observed in \citet{More2015} is recovered, but only for halos with masses greater than 5$\times$10$^{12}$M$_\odot$. Lower mass objects show a less smooth evolution of the splashback radius with $\Gamma$. 

Furthermore, the mean value of the splashback radii calculated from individual halos produces very different results than the splashback radii of the stacked samples, as can be seen in Fig. \ref{Fig:splashtest}.  This suggests that the density distribution of the different halos at a given radius is not Gaussian,  and the  two different definitions of the mean are not equivalent. This is beyond the scope of this paper but would benefit from further exploration. Alternatively, the different approaches used in the analysis simply provide different estimates of these quantities.  By assuming that the splashback radius is the deepest dip in the density gradient profile, we would confuse the dip due to the second orbit with the true splashback radius further out \citep{Adhikari2014}.

For the splashback radii of individual halos we use a subselection of halos from Illustris-1-Dark. We calculate the splashback radii of the density gradient profiles from 0-5 R$_{vir}$ and 0.5 to 2.5 R$_{vir}$ separately, and use only those halos which give the same value. This was to ensure that the calculation is reasonably robust given halo-to-halo noise. This same procedure was used for the splashback radii presented in the main text as well. 

If we use the median profiles, as in \citet{Diemer2014} the two caustic feature is still apparent, but not as significant. The size of the outer shell is reduced in the whole sample but remains important, especially in higher mass bins.

\section{Resolution effects}
\label{Append_res}

\begin{figure}
\centering
\includegraphics[scale=.4, trim={0.cm 0 0cm 0cm},clip]{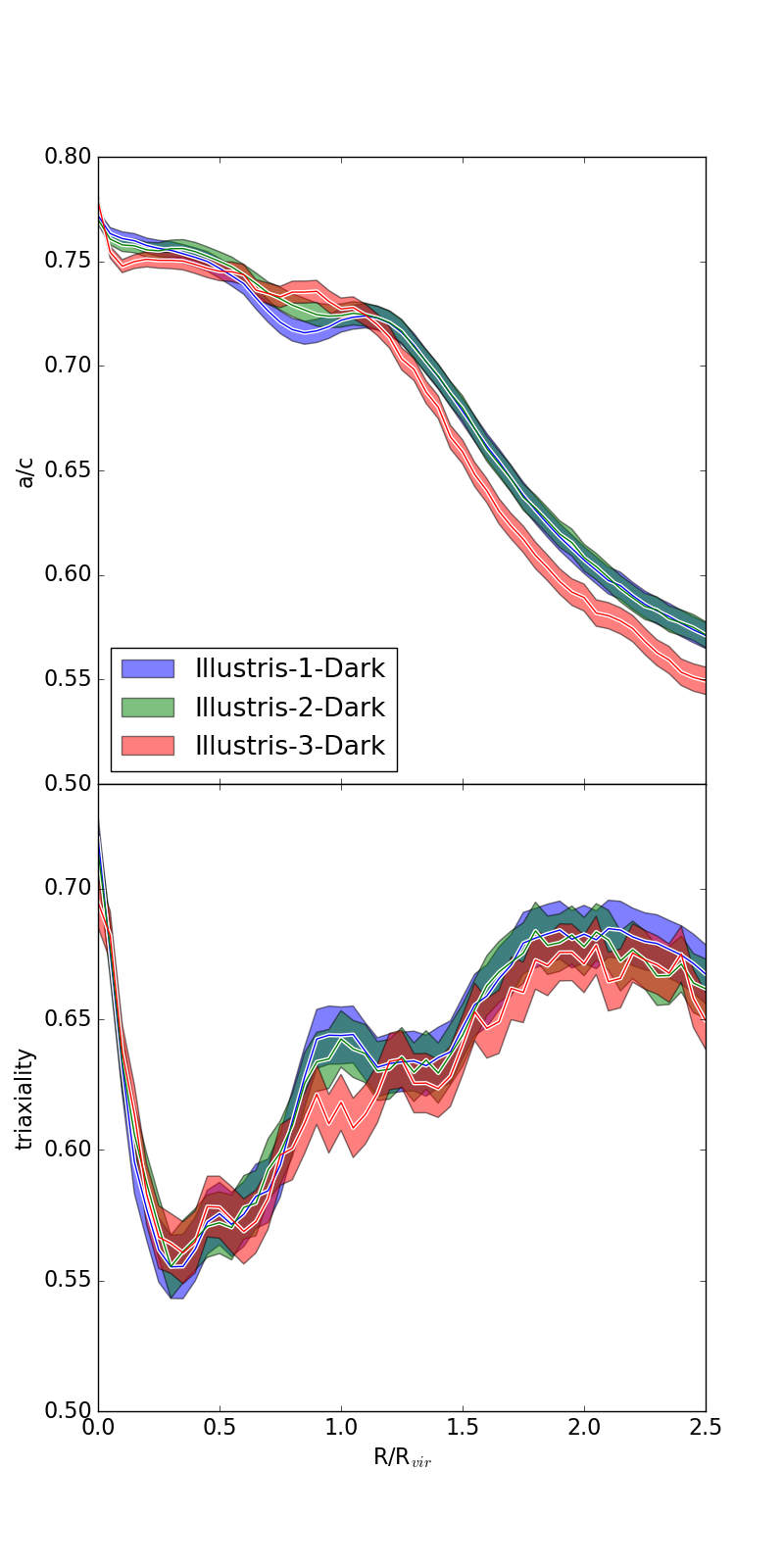}
\caption{The sphericity (top panel) and triaxiality (bottom panel) for the three Illustris-Dark simulations. The mass range for M$> 8.5\times 10^{12} M_\odot$ in each simulation. Only the lowest resolution simulations does not show a clear feature close to the virial radius. }
\label{Fig:shaperes}
\end{figure}

Figure \ref{Fig:shaperes} shows that the shape features in the mean halos from the different Illustris samples vary depending on the resolution of the simulation. We use the first 300 halos from high resolution Illustris-1-Dark sample and 300 most massive halos from the Illustris-2-Dark sample along with the 300 most massive halos from Illustris-3-Dark.  

 The key shape feature in both the halo sphericity and triaxiality is missing from the lowest resolution Illustris-3-Dark sample. This is because the number of particles inside each radial bin drops by a factor of eight between Illustris-2-Dark and Illustris-3-Dark, and by a factor of 64 between Illustris-1-Dark and Illustris-3-Dark. There is much more noise in the shape profile in Illustris-3-Dark which acts to hide the feature as the resolution decreases.

It is evident in the upper panel of Fig. \ref{Fig:shaperes} that the bump feature, in both sphericity and triaxiality, becomes stronger with increased resolution. The two distributions overlap at high and low radius, but the drop in sphericity close to the virial radius is even stronger in the higher resolution sample, emphasizing the importance of the feature.

The lower panel of Fig. \ref{Fig:shaperes} presents the mean triaxiality parameter with radius, scaled by the virial radius, as before. The  mean triaxiality profile falls with radius out to half the virial radius, then starts to increase again, becoming more prolate at large `r'. The high and medium resolution cases contain a peak at around the virial radius which is missing at lower resolution. This implies that information in the structure of the dark matter is being lost at lower resolution. The mean halo in our Illustris-1-Dark sample is prolate at the centre, triaxial at half $R_{vir}$ and becomes prolate again at larger distance, {\it with a strongly prolate feature close to the virial radius.} This implies that the halos know about their virial radius {\it on average}.

\section{MUGS and Illustris profiles and the effect of substructure}
\label{App:MugIl}

\begin{figure}
    \centering
     \includegraphics[scale=.4]{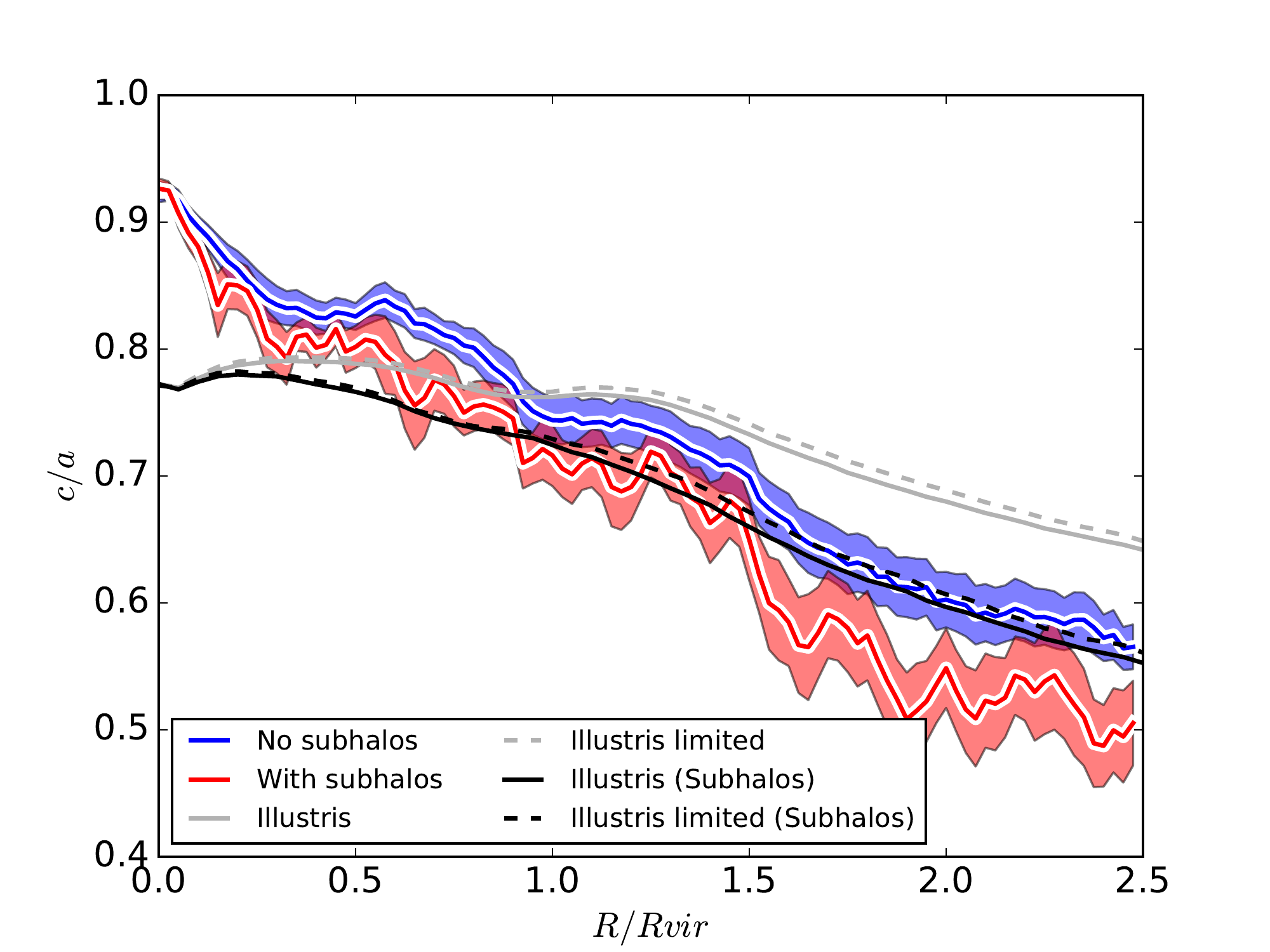} 
     \caption{The sphericity profile of the mean MUGS halo with (red) and without (blue) subhalos included. The shaded area is the standard error on the mean. The solid Illustris curves shows the mean profile of all the Illustris-1-Dark sample with (black) and without subhalos (grey).The dashed lines show the same for a subsample of halos with virial mass between 1.6$\times 10^{12}$ and 3.5$\times 10^{11}$, which is the mass range in MUGS.}
     \label{Fig:MeanMUGSz0}    
\end{figure}

Figure \ref{Fig:MeanMUGSz0} shows the average shape profile of the 16 halos in MUGS with and without subhalos. Also included is the average Illustris-1-Dark sphericity profile, with and without subhalos.  

 There is a noticeable difference in the sphericities of halos in MUGS and Illustris-1-Dark. The inner regions of halos in MUGS are significantly more spherical than those in Illustris.  The first possible explanation is that there is a selection effect in MUGS that favours halos with more spherical inner regions, such as the mass range etc.  \citet{Jeeson2011} observe a correlation between halo sphericity and mass using a Spearman rank co-efficient on widely scattered data. Alternatively, the difference in the inner regions may be due to the hydrodynamical processes included in MUGS, which acts to sphericalise the halos. We have  used the dark matter only version of Illustris because we are interested in the halo outskirts, which are less affected by baryons.

The mean sphericity profile for MUGS (Fig. \ref{Fig:MeanMUGSz0}, blue region) exhibits a marked flattening between 1 and 1.2 times the virial radius, and another at 0.5 times the virial radius. The inner change of gradient may mark the region where the baryons become important. When subhalo particles are included in the shape profile calculation these features cannot be seen due to the increased noise in the distribution. Further, the shape profile in the outer region of the figure (beyond 1.5R$_{vir}$) are more spherical in Illustris-1-Dark compared to MUGS  for the same mass range. This cannot be put down to a resolution difference as Fig. \ref{Fig:shaperes} shows that the inner profiles are unaffected by resolution (see Appendix \ref{Append_res} for details). Of course, the MUGS sample selection contains only 16 galaxies. In 5\% of subsamples of 16 halos from Illustris-1-Dark (in the same mass range) the sphericity value of 0.6 at 2.5 R$_{vir}$ can be reached, but in no subsample is a sphericity of 0.9 at the halo centre achieved. The outer regions of halos in MUGS are also indicative of much higher mass halos in Illustris.

We examined a massive halo from the Illustris-2-Dark and the same object in Illustris-2 simulation in order to explore the effect of substructure and baryons. The inner regions of this halo are more spherical when baryons are included, and almost identical further out. The impact of subhalos and halos in the local environment can also be seen, and it is evident that these (sub)halos increase the noise in the profile, but keep the same overall profile. The lack of baryons in N-body simulations, such as Illustris-2-Dark, result in less spherical halos, especially in the inner regions, although the effect is limited to inside 0.5R$_{vir}$. However, the effect appears to be on the same order as would be required to increase the sphericity in the inner regions of halos close to the value for MUGS.

We can carry out a similar analysis on the MUGS sample. By comparing g15784 with simulations carried out using both different baryonic physics  (the MaGICC run), and a separate simulation using only dark matter particles, but the same initial conditions, we find that the dip is equally prominent in MaGICC, the dark matter-only simulation, and the MUGS version.  There is some difference in the final form of sphericity profile between the runs with and without baryons, but most of this appears to be in the inner regions, as found by numerous other authors \citep[e.g.][]{Bailin2005}, rather than near the virial radius, where the potential is dominated by the dark matter.

\begin{figure}
\centering
\includegraphics[scale=.4, trim={0.cm 0 0cm 0cm},clip]{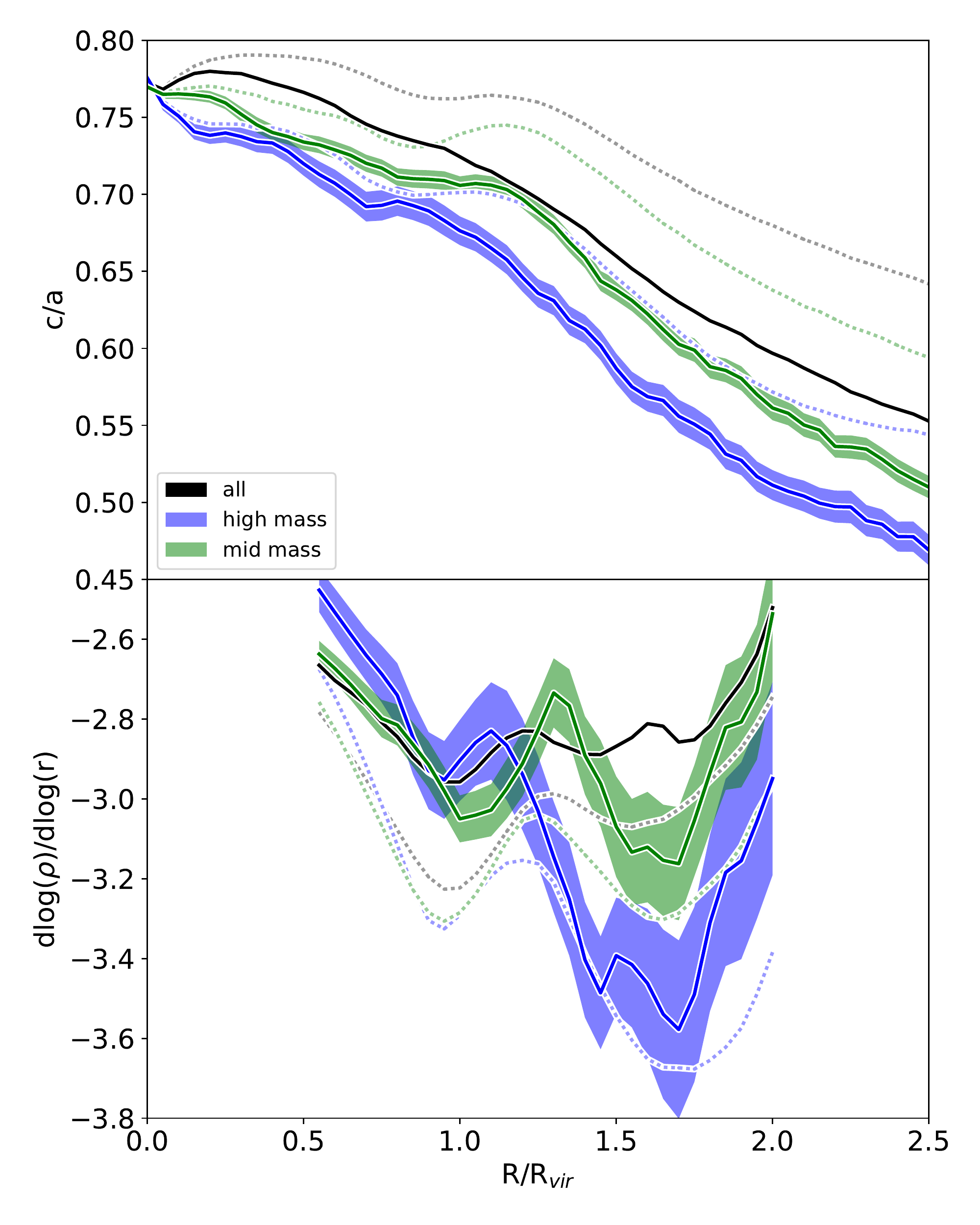}
\caption{The mean sphericity and density gradient profile of dark matter halos extracted from the Illustris-1-Dark sample {\it including} subhalos. The different lines show different subsamples, while the shaded regions are the corresponding standard error ($\sigma/\sqrt{N}$). {The upper panel shows the sphericity gradients and the lower panel shows the density gradient profiles. The gradient profiles are only shown between 0.5 R$_{vir}$ and 2 R$_{vir}$. The dotted lines show the profiles without subhalos from Fig. \ref{Fig:Averages}}.}
\label{Fig:shapesplashSH}
\end{figure}

\begin{figure}
\centering
\includegraphics[scale=.4, trim={0.cm 0 0cm 0cm},clip]{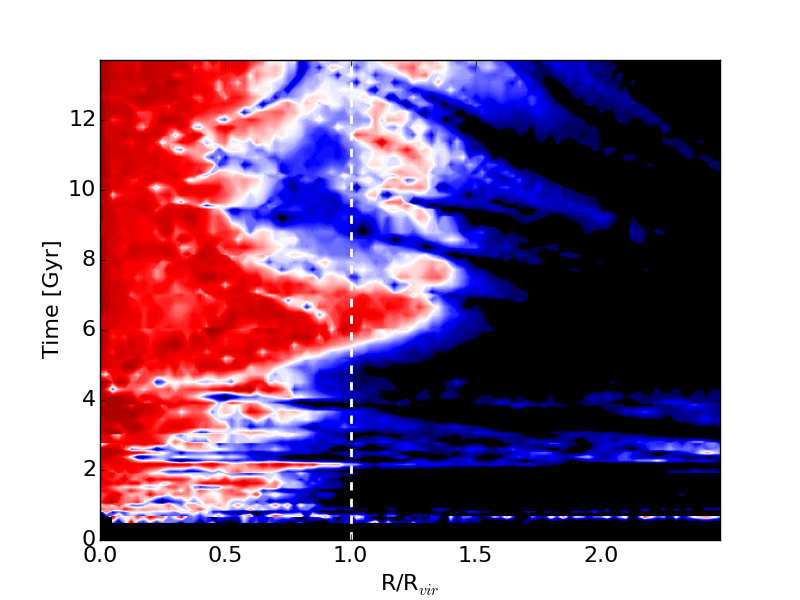}
\caption{The sphericity profiles of the MUGS galaxy g4720 including substructure. Colour scheme as in Fig. \ref{Fig:g15784zrevolsph}, \ref{Fig:sphfeaturemugs} and \ref{Fig:sphfeaturemugs2}}
\label{Fig:examplewithsh}
\end{figure}

Where subhalos are included, see Figs. \ref{Fig:MeanMUGSz0}, \ref{Fig:shapesplashSH} and \ref{Fig:examplewithsh}, the results become significantly noisier, and the bin to bin scatter is considerably stronger in the density gradient profile as well. The  two caustics in the splashback radius seen in Fig. \ref{Fig:Averages} is still apparent in Fig. \ref{Fig:shapesplashSH}, but the sphericity feature is lost (except in the highest mass bin). Clearly, the calculation of the profiles with substructure makes the details of the profiles more difficult to identify, potentially requiring larger samples to recover or are simply lost in the noise. In the bottom panel of Fig. \ref{Fig:shapesplashSH} we compare different density gradient profiles to each other \citep[e.g.][]{Diemer2014}, in this case the profiles with and without substructure in different mass bins.  We see the following effects caused by including substructure in the density gradient profile: (1) The depth of the local minima at $R_{vir}$ and $\sim$1.7$R_{vir}$ are shallower, and the density gradient is, overall, less steep compared to the density gradient excluding substructure, (2) the density gradient profile is less smooth than in Fig. \ref{Fig:Averages}, which is a result of the increased noise introduced by the substructure, (3) in the entire sample (the black line) the minima at 1.7$R_{vir}$ cannot be identified when substructure is included. Point (3), however, is only true for the whole sample, and is a result of the sensitivity of the density gradient in low mass halos to substructure. This demonstrates the importance of removing substructure in order to probe the shape of the background density field.

In MUGS, the virial shape features are more difficult to identify if substructure is retained, even with the time evolution data (Fig. \ref{Fig:examplewithsh}). The substructure in appears as thin marks superimposed on the background distribution, disguising the underlying halo shape evolution. With the substructure removed, the underlying profiles are clearer, but are shown not to be the result of the substructure removal method. 

Removing substructure, however, may, of course, introduce uncertainties due to the halo finder, and how it extracts substructure from the host. This discussion is beyond the scope of the paper and should be addressed in future work.

\end{document}